\documentstyle[prc,preprint,float,aps,epsfig,amsmath,rotating]{revtex}

\newboolean{intro}
\setboolean{intro}{true}

\begin{document}

\tighten

\title{Double radiative pion capture on hydrogen and deuterium 
and the nucleon's pion cloud}
\author{S. Tripathi $^1$, D.S. Armstrong $^2$, 
M.E.~Christy $^1$, J.H.D.~Clark,$^{4}$\cite{APSaddress},
T.P. Gorringe $^1$, M.D. Hasinoff$^3$, 
M.A. Kovash$^1$, D.H.~Wright,$^{3}$\cite{SLACaddress},
and P.A.~{\.Z}o{\l}nierczuk$^{1}$.}

\vspace{4.0cm}

\address{$^1$ Department of Physics and Astronomy,
University of Kentucky,\\ Lexington, KY, USA 40506.} 

\address{$^2$ Department of Physics, College of William and Mary,
\\ Williamsburg, VA, USA 23187.}

\address{$^3$ Department of Physics and Astronomy, University of
British Columbia, \\ Vancouver, BC, Canada V6T 1Z1.}

\date{\today}

\maketitle


\abstract
{We report measurements of double radiative capture 
in pionic hydrogen and pionic deuterium.
The measurements were performed with the RMC spectrometer 
at the TRIUMF cyclotron by recording photon pairs 
from pion stops in liquid hydrogen and deuterium targets.
We obtained absolute branching ratios
of $( 3.02 \pm 0.27 (stat.) \pm 0.31  
(syst.) ) \times 10^{-5}$ for hydrogen
and $( 1.42 \pm ^{0.09}_{0.12} (stat.) \pm 0.11  
(syst.) ) \times 10^{-5}$ for deuterium, and 
relative branching ratios 
of double radiative capture  to single radiative capture
of $( 7.68 \pm 0.69(stat.) \pm 0.79(syst.)) \times 10^{-5}$ 
for hydrogen and $( 5.44 \pm^{0.34}_{0.46}(stat.) 
\pm 0.42(syst.)) \times 10^{-5}$ for deuterium.
For hydrogen, the measured branching ratio
and photon energy-angle distributions
are in fair agreement with a 
reaction mechanism involving the annihilation of
the incident $\pi^-$
on the $\pi^+$ cloud
of the target proton.
For deuterium, the measured branching ratio
and energy-angle distributions
are qualitatively consistent
with simple arguments
for the expected role of the spectator neutron.
A comparison between our hydrogen and deuterium data
and earlier beryllium and carbon data
reveals substantial changes 
in the relative branching ratios
and the energy-angle distributions
and is in agreement with the expected evolution of the reaction dynamics
from an annihilation process in S-state capture 
to a bremsstrahlung process in P-state capture.
Lastly, we comment on the relevance of the double radiative process
to the investigation of the charged pion polarizability 
and the in-medium pion field.}


\section{Introduction}
\label{s-introduction}


A cornerstone of quantum chromodynamics at low energies 
is the pion's emergence
as the approximate Goldstone boson
of a spontaneously broken 
chiral symmetry.
Consequently, the pion is central to understanding 
the realization of QCD symmetries
and their associated currents 
in low energy hadronic processes. 
In particular, the emission and absorption 
of pions by nucleons is closely connected
to the low energy manifestation of the 
partially conserved axial current.


Radiative pion capture, $\pi N \rightarrow \gamma N$,
and pion photo-production,  $\gamma N \rightarrow \pi N$,
are well-known benchmarks 
for experimental tests of the low energy realization
of approximate chiral symmetry.
The chiral predictions for pion photo-production
and radiative pion capture
were originally obtained from current algebra techniques \cite{Kr54}
and are nowadays derived from heavy 
baryon chiral perturbation theory \cite{Be96,Fe00}.
Such arguments relate 
the threshold amplitude for absorption or emission
of s-wave pions to the axial currents
of the nucleon 
and the pion. 
The resulting relationship 
for $\gamma p \rightarrow \pi^+ n$ photo-production
-- the Kroll-Ruderman theorem
and higher-order corrections -- 
is experimentally verified to a few-percent accuracy.


Less well known -- both theoretically and experimentally --
is double radiative pion capture, $\pi N \rightarrow \gamma \gamma N$,
and radiative pion photo-production, $\gamma N \rightarrow \gamma \pi N$.
Nonetheless the absorption or emission of pions
in double radiative capture or radiative photo-production 
is similarly constrained by chiral symmetry.
Predictions for double radiative pion capture
$\pi^- p \rightarrow \gamma \gamma n$
have been derived from a current algebra approach
by Lapidus and Muskhanov \cite{La72}
and a chiral Lagrangian approach
by  Beder \cite{Be79,Be82} and Cammarata \cite{Ca79}.
Most interestingly, 
all studies have suggested
that threshold absorption
from $\ell=0$ orbitals involves 
the annihilation
of the incident $\pi^-$
on the $\pi^+$ cloud 
of the target proton,
the double radiative process thus exhibiting intriguing sensitivity
to the nucleon's pion cloud
in the small-to-moderate $q^2$ regime.


Originally,
some excitement was generated 
by the prediction
of $\pi^- \pi^+ \rightarrow \gamma \gamma$ annihilation
in double radiative capture.
The context was speculations on pion condensation 
or precursor effects 
in ordinary nuclei,
which motivated Ericson and Wilkin \cite{Er75}, 
Nyman and Rho \cite{Ny77},
and Barshay \cite{Ba78}
to advance the $( \pi , 2\gamma )$ reaction
on nuclear targets
as a selective probe 
of the pion field in the nuclear medium. 
While pion condensation at nuclear densities is 
nowadays unimaginable, the investigation of the 
pion field in the nuclear medium
is still of continuing importance.


Recently, new studies \cite{Ah05}
have attempted to determine the pion's electromagnetic polarizability
via radiative pion photo-production.
The method involves the Compton scattering
of the incident $\gamma$-ray
from the $\pi^+$ cloud
of the target proton
-- {\it i.e.}\ the proton acting
as a pion target -- with the polarizability being 
encoded in the amplitude for the
$\gamma \pi \rightarrow \gamma \pi$ scattering.
Consequently, a better understanding of the pion cloud
contribution to $\gamma \pi \rightarrow \gamma \pi$ 
scattering in radiative photo-production
or $\pi \pi \rightarrow \gamma \gamma$ annihilation 
in double radiative capture
is of some importance.


Unfortunately, 
the predicted branching ratios for 
double radiative capture in pionic atoms are small 
-- of order of $10^{-4}$ -- and the experimental work 
is extremely challenging.
Nevertheless, experiments 
by Deutsch {\it et al.}\ at CERN \cite{De79}
and Mazzucato {\it et al.}\ at TRIUMF \cite{Ma80}
were successful in recording gamma-ray pairs 
from double radiative capture 
on $^9$Be and $^{12}$C targets. 
However, the elementary process
of double radiative pion capture 
on hydrogen, and the simplest nuclear process
of double radiative pion capture 
on deuterium, have so far remained unobserved.\footnote{A
branching ratio upper limit $\leq$ 5.5$\times$10$^{-4}$
for double radiative capture on pionic hydrogen was obtained by
Vasilevsky {\it et al.}\ \cite{Va69}.}


Herein we report the results of experimental work
on double radiative capture in pionic hydrogen 
and pionic deuterium.
The experiment was conducted with the RMC spectrometer
at the TRIUMF cyclotron and detected photon-pairs
from pion stops in liquid hydrogen and deuterium targets.
Our results include absolute branching ratios,
relative branching ratios,
and photon energy-angle distributions
for double radiative capture on
hydrogen and deuterium.
Our goals in conducting the measurements were to:
test the predictions 
for the reaction dynamics
of the elementary process,
compare the results 
for the elementary process
and the simplest nuclear process,
and examine the evolution of
double radiative capture from 
$Z = 1$ to complex nuclei. A letter on our experimental results
for the elementary process
has been previously published \cite{Tr02}.\footnote{In addition,
in Refs.\ \cite{Zo04,Zo02} we conducted searches for 
exotic baryon and dibaryon resonances
by detection of photon-pairs following pion capture 
in hydrogen and deuterium.}

This article is organized as follows. 
In Sec.\ \ref{theory} we briefly outline the theoretical aspects 
of the  $\pi N \rightarrow \gamma \gamma N$ process.
In Sec.\ \ref{backgrounds} we discuss the 
background difficulties arising
from real $\gamma$-ray coincidences
and accidental $\gamma$-ray coincidences 
from  pion charge exchange, single radiative capture 
and other processes.
Sec.\ \ref{setup} describes the experimental setup
and trigger electronics.
The identification of the double radiative capture events, 
subtraction of the two-photon background events,
and determination of the two-photon detection efficiency 
are described in Sec.\ \ref{analysis}.
In Sec.\ \ref{results}
we discuss our results for hydrogen and deuterium,
and compare our experimental data for these elementary processes
with earlier data for the nuclear process.
Our conclusions 
are presented in Sec.\ \ref{conclusions}.

\section{Theoretical overview}
\label{theory}

\subsection{Initial state}

Pionic atoms are initially formed
in atomic orbitals with 
large principal quantum numbers ($n$)
and an approximately statistical population
of the orbital sub-states ($\ell$).
The atoms then de-excite by a combination
of Auger transitions 
and radiative transitions 
with the cascade ultimately terminating 
in the nuclear absorption process (for details see Ref.\ \cite{Ba82}). 

In pionic hydrogen and pionic deuterium 
the nuclear absorption is dominated
by S-state capture from $n = 3-6$ orbitals \cite{Da60,Le62,Bo80,Te97}.
This unique feature of $Z = 1$ capture originates
from the combination of strong Stark mixing 
between degenerate $\ell$-orbitals
and weak $\ell > 0$ nuclear absorption
and is firmly established
by the available pionic x-ray and strong 
interaction data \cite{Da60,Le62,Bo80,Te97,Br51}.

By comparison, in pionic beryllium and pionic carbon
the nuclear capture is predominantly from P-states.  
For $\pi^-$$^{12}$C, Sapp {\it et al.}\ \cite{Sa72} have given
summed absorption probabilities of $0.08\pm 0.03$ for S-state capture,
$0.92\pm 0.03$ for P-state capture, and $\sim$$7 \times 10^{-4}$ for  
D-state capture. For $\pi^-$$^{9}$Be, Sapp {\it et al.}\ \cite{Sa72} 
have tabulated x-ray yields that imply summed absorption probabilities 
of $\sim$0.2 for S-state capture
and  $\sim 0.8$ for P-state capture.

\subsection{Reaction kinematics}

\begin{figure}
\begin{center}
\epsfig{file=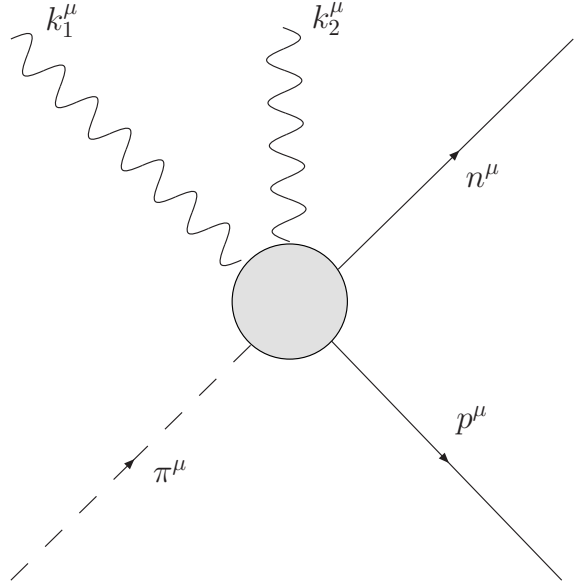,height=8.0cm}
\end{center}
\caption{Definitions of kinematical quantities
in the $\pi^- p \rightarrow \gamma \gamma n$ 
elementary process.}
\label{fig:kinematics}
\end{figure}

For the  $\pi^- p \rightarrow \gamma \gamma n$ elementary process
the important kinematical quantities  
are shown in Fig.\ \ref{fig:kinematics}. 
We denote the pion 4-momentum $\pi^{\mu} = ( m_{\pi} , 0 )$, 
the proton 4-momentum $p^{\mu} = ( M_p , 0 )$, 
the neutron 4-momentum $n^{\mu} 
= ( E_n , \vec{p_n} )$, 
and the photon 4-momenta as $k_1^{\mu} = ( \omega_1 , \vec{k_1} )$ 
and $k_2^{\mu} = ( \omega_2 , \vec{k_2})$.
The pion, proton and neutron masses 
are $m_{\pi}$, $M_p$ and $M_n$  respectively.

The measured quantities in our experiment
are the two photon energies $\omega_1$, $\omega_2$ 
and the photon-pair opening angle cosine, $y = \cos{\theta}$. 
In terms of measured quantities
the 3-momentum transfer $\vec{q} = \vec{p_n}$
and energy transfer $q_o = E_n - M_p$ are
\begin{equation}
| \vec{q} |  =  \sqrt{\omega_1^2 + \omega_2^2 + 2 \omega_1 \omega_2 y}
\end{equation}
and
\begin{equation}
q_o  =  m_{\pi} - \omega_1 - \omega_2 \\
\end{equation}
where the binding
energy of the pionic atom has been neglected.
Note that the 4-momentum transfer-squared $q^2$
is space-like and 
ranges from zero,
for back-to-back photons with equal energy partition,
to -$m_{\pi}^2$,
for either parallel photons or a photon energy at the 
end-point energy.

In the $\pi^- p \rightarrow \gamma \gamma n$ elementary process
the photon energies range from $0$ to $129.4$~MeV
and the neutron kinetic energy ranges  from $0$ to $8.9$~MeV.
The $\gamma \gamma n$ 3-body phase space distribution
strongly favors the equal sharing of the available
energy between the two photons and
somewhat favors large photon-pair opening angles over 
small photon-pair opening angles.

\subsection{Reaction dynamics}
\label{dynamics}

Calculations of the 
$\pi^- p \rightarrow \gamma \gamma n$ elementary process 
have been performed 
by Joseph \cite{Jo60}, 
Lapidus and Musakanov \cite{La72} 
and Beder \cite{Be79} for at-rest capture
and Beder \cite{Be82}
and Cammarata \cite{Ca79}
for in-flight capture.
For at-rest capture on hydrogen the 
predicted absolute branching ratios 
are 5.1$\times$10$^{-5}$ \cite{Jo60},
5.1$\times$10$^{-5}$ \cite{La72}
and 5.4$\times$10$^{-5}$ \cite{Be79},
and the corresponding ratios $R( 2\gamma  / 1\gamma )$
of double radiative capture
to single radiative capture
are  $1.4 \times 10^{-4}$,
$1.4 \times 10^{-4}$
and $1.3 \times 10^{-4}$, respectively.

\begin{figure}
\begin{center}
\epsfig{file=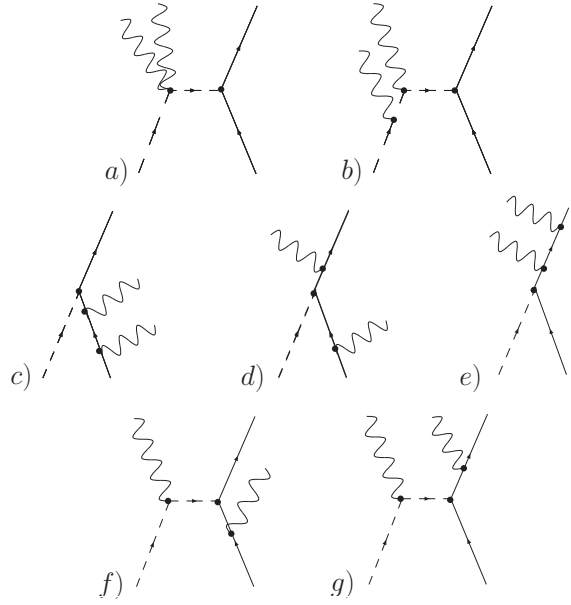,height=8.0cm}
\end{center}
\caption{Tree-level contributions to double
pion radiative capture in the $\pi N \gamma$ effective 
Lagrangian approach of Beder. Diagrams
a-b) are denoted the $\pi$$\pi$  graphs,
c-e) the NN graphs,
and f-g) $\pi$N graphs.}
\label{fig:feydiag}
\end{figure}

At tree-level the $\pi N \gamma$ effective Lagrangian
of Beder \cite{Be79} implies three categories of
contributions to double radiative capture:
the $\pi \pi$ diagrams
of Fig.\ \ref{fig:feydiag}.a-b, with both photons
attached to the pion,
the NN diagrams
of Fig.\ \ref{fig:feydiag}.c-e, 
with both photons attached to the nucleon,
and the $\pi N$ diagrams
of Fig.\ \ref{fig:feydiag}.f-g, with one photon 
attached to the pion and one photon attached to the nucleon.
The graphs in  Fig.\ \ref{fig:feydiag}.a-b
and \ref{fig:feydiag}.f-g are leading order $O(1)$,
and the graphs in Fig.\ \ref{fig:feydiag}.c-e
are next-to-leading order $O(1/M_n)$,
in the modern terminology of heavy-baryon chiral
perturbation theory \cite{Ka04}.

In threshold $\ell = 0$ capture 
({\it e.g.}\ $^1$H and $^2$H)
the contributions of diagrams \ref{fig:feydiag}.b and \ref{fig:feydiag}.f-g 
vanish and the only surviving leading-order graph 
is the four-point $\pi$$\pi$ annihilation 
graph of Fig.\ \ref{fig:feydiag}.a. 
The four-point $\pi \pi$ annihilation diagram 
yields a squared matrix element 
with kinematical dependence \cite{Be79}
\begin{equation}
| M.E.|^2 \propto (\omega_1^2 + \omega_2^2 + 2 \omega_1 \omega_2 y )
( 1 + y^2 ) /
( m_{\pi} (\omega_1 + \omega_2) - \omega_1 \omega_2 (1 - y) )^{2}
\label{pipi}
\end{equation}
that peaks at small opening angles
and equal energy partition.
This sole $O(1)$ contribution to $\ell = 0$ capture,
involving annihilation of the stopped $\pi^-$ on
the $\pi^+$-field of the target proton,
and offering sensitivity to the nucleon's
pion cloud and the $\pi\pi \rightarrow \gamma\gamma$ vertex,
is the most intriguing feature of the 
double radiative process.

In addition,  
the NN diagrams 
of \ref{fig:feydiag}.c-e introduce
a next-to-leading order $O(1/M_n)$ contribution
to threshold $\ell = 0$ capture.
These graphs yield a squared matrix element 
with kinematical dependence \cite{Be79} 
\begin{equation}
|M.E.|^2 \propto ( 1 - y )^2
\label{ngraph}
\end{equation} 
that peaks
at large opening angles
but is independent
of the energy partition
between the two photons.
According to Beder \cite{Be79},
the $O(1)$ $\pi$$\pi$ graph \ref{fig:feydiag}.a
contributes roughly 65\%,
the $O(1/M_n)$ NN graphs \ref{fig:feydiag}.c-e
contribute roughly 20\%,
and their interference
terms contribute roughly 15\%,
to the $\ell = 0$ total branching ratio.

In threshold $\ell = 1$ capture 
({\it e.g.}\ $^{9}$Be and $^{12}$C)
the leading order $\pi$$N$ diagrams \ref{fig:feydiag}.f-g 
are found to dominate,
with small contributions from the $O(1)$ $\pi$$\pi$ 
graphs and negligible contributions from the $O(1/M_n)$ $N$$N$ graphs.
The $\pi$$N$ bremsstrahlung graphs \ref{fig:feydiag}.f-g
yield a squared matrix element 
with kinematical dependence \cite{Be79}
\begin{equation}
|M.E.|^2 \propto 
1 / \omega_1^{2} + 1 / \omega_2^{2} + ( 1 + y^2 ) / (2 \omega_1 \omega_2 ) 
\label{}
\end{equation}
that diverges for small photon energies.
A striking difference between
$\ell=0$ capture and $\ell=1$ capture
is the prediction of a roughly ten-fold increase in 
the ratio $R( 2\gamma / 1\gamma )$.

In addition
to the above
contributions from the $\pi N \gamma$
effective Lagrangian, the possible effects
of virtual $\pi^o$'s and intermediate $\Delta$'s 
have been investigated 
by several authors \cite{La72,Er75,Ba78}.
The contribution of virtual $\pi^o$'s,
which involves virtual $\pi^{\circ \star} \rightarrow \gamma \gamma$ decay
following off-shell $\pi^- p \rightarrow \pi^{\circ \star} n$ charge exchange,
was found by Lapidus and Musakanov \cite{La72},
Ericson and Wilkin \cite{Er75}
and Barshay \cite{Ba78}
to be entirely negligible.
The contribution of virtual $\Delta$'s,
which involve $\Delta$ excitation by pion absorption and
$\Delta$ decay by photon emission,
was estimated by Beder \cite{Be79} 
to be roughly 5\% in the $\ell=0$ process.

\subsection{Medium effects}

Of course, in double radiative capture on nuclear
targets the dynamics of the elementary process
must be folded with the nuclear response.
Although no calculation for double radiative capture 
on deuterium has been performed a number of studies of double
radiative capture on light nuclei have been published.
The most detailed treatments of nuclear double
radiative capture were published by Christillin and Ericson \cite{Ch79} and
Gil and Oset \cite{Gi95}.

Christillin and Ericson \cite{Ch79} combined an 
effective interaction incorporating
both $\pi$$\pi$ and $\pi$N graphs
with a phenomenological nuclear resonance
incorporating the giant dipole and other 
resonances to calculate $^{12}$C double radiative 
capture. They found that an important effect
of the nuclear medium was a substantial quenching
of the partial rate in the kinematical region 
$( \vec{k_1} +\vec{k_2} ) \rightarrow 0$ due to Pauli blocking.
Gil and Oset \cite{Gi95} combined an 
effective interaction that includes
the $\pi$$\pi$ graphs but omits
$\pi$N graphs with the Fermi gas model and
the local density approximation 
to calculate $^{9}$Be and $^{12}$C double radiative capture.
In contrast to Christillin and Ericson
they found a significant enhancement
of the partial rate in the kinematical region $( \vec{k_1} +\vec{k_2} ) 
\rightarrow 0$, and commented that medium modification
of the pion field has a large effect for these kinematics.
Note that both calculations indicated a large increase 
in the ratio R(2$\gamma$/1$\gamma$) for the nuclear
process over the elementary process.
 
\section{Experimental challenges}
\label{backgrounds}

The measured values of the branching ratios 
for the known modes of pion absorption on hydrogen and deuterium
are listed in Table \ref{t: H BR}.
For hydrogen the capture modes
are pion charge exchange, $\pi^- p \rightarrow \pi^o n$,
single radiative capture, $\pi^- p \rightarrow \gamma n$,
and radiative pair production, $\pi^- p \rightarrow e^+ e^- n$
\cite{Sp77,Sa61,Fo89}.
For deuterium the capture modes are 
non-radiative capture $\pi^- d \rightarrow n n$,
radiative capture $\pi^- d \rightarrow \gamma n n$,
and pion charge exchange $\pi^- d \rightarrow \pi^o n n$ \cite{Hi81,Ma77}.
By comparison the anticipated branching ratios for double radiative capture
on hydrogen and deuterium are of order $10^{-4}$ --
the tiny branching ratios,
neutral particle final-states
and continuum energy-angle distributions
together making detection quite difficult.
Moreover, these difficulties are magnified
by intense backgrounds from real coincidences and accidental coincidences
of photon pairs.

Neutral pion decay $\pi^o \rightarrow \gamma \gamma$
following at-rest pion charge exchange 
yields real $\gamma$-$\gamma$ coincidences,
the $\pi^o$ recoil energy
of $2.9$~MeV in $\pi^- p \rightarrow \pi^o n$
and $0-1$~MeV in $\pi^- d \rightarrow \pi^o n n$
imparting a Doppler broadening on the 
laboratory energy-angle distributions
of the $\pi^o \rightarrow \gamma \gamma$ photons
(the two-photon background from in-flight charge exchange
was negligible compared to at-rest charge exchange).
In hydrogen the broadening yields
an energy spread $55.0 < \omega < 83.0$~MeV
and an angular spread $-1 < y < -0.91$.
In deuterium the broadening yields
an energy spread $61.6 < \omega < 73.9$~MeV
and an angular spread $-1 < y < -0.97$.
The $\pi^o \rightarrow \gamma \gamma $ background 
therefore prohibits a measurement of double radiative capture 
at large opening angles
and endangers the measurement at small opening angles 
due to the finite resolution of the detector system.
Since the branching ratio for charge exchange
is much larger in hydrogen ($0.607 \pm 0.004$ \cite{Sp77}) than deuterium 
($( 1.45 \pm 0.19 ) \times 10^{-4}$ \cite{Hi81})
the $\pi^o \rightarrow \gamma \gamma $ background 
is more worrisome in hydrogen than in deuterium.
Herein we denote the real $\gamma$-$\gamma$ coincidences 
from $\pi^o \rightarrow \gamma \gamma$ decay 
as the ``$\pi^o$ background''.

Charge exchange and single radiative capture yield photon-pairs
via accidental $\gamma$$\gamma$ coincidences.
In our experiment the incident beam had a micro-structure with a pulse
period of 43~ns and a pulse duration of 2$-$4~ns.
Consequently, 
the probability of two pions stops 
in one beam pulse
was not negligible.
Such multiple pion stops can
generate accidental $\gamma$-$\gamma$ coincidences
via the detection of one photon from 
one pion stop and another photon
from another pion stop
in the same beam pulse.
These accidental $\gamma$-$\gamma$ coincidences
may comprise a pair of photons that originate
from: two charge exchange events, two single radiative capture events,
or one charge exchange event and one single radiative capture event.
Such accidental $\gamma$-$\gamma$ coincidences
have summed energies $106 < \omega < 258$~MeV
and opening angles $-1 < y < +1$.
Herein we denote the accidental $\gamma$-$\gamma$ coincidences 
from multiple pion stops in one beam pulse 
as the ``$2$$\pi$ background''.

Lastly, two possible sources of two-photon backgrounds
were (i)  Dalitz decay following pion charge exchange
and (ii) $\mu^-$ decay from in-beam muon contamination.
In background process (i) the decay $\pi^o \rightarrow \gamma e^+ e^-$
can yield a photon-pair by identification of one
true photon (the Dalitz-decay $\gamma$-ray) 
and one fake photon (the Dalitz-decay $e^+$$e^-$ pair).
In background process (ii) the $\mu^-$ decay yields
single photons via internal bremsstrahlung or external bremsstrahlung
and photon pairs by accidental coincidences with
other photons from pion stops.
Although both backgrounds are much rarer than
the $\pi^o$ background or the $2$$\pi$ background,
they are less straightforwardly differentiated from the double radiative
capture events by the applied cuts (see Sec.\ \ref{signal events}). 
Herein we denote background (i) as the Dalitz background
and background (ii) as the $\mu$$\pi$ background.

\section{Experimental Setup}
\label{setup}

Our  measurements of $( \pi^- , 2\gamma )$ capture
were conducted with the
RMC detector on the M9A beamline 
at the TRIUMF cyclotron. 
The RMC detector  is a 
photon-pair spectrometer
with large solid angle coverage 
and moderate energy resolution.
The detector was developed for studies
of radiative muon capture on hydrogen and nuclei,
and has been discussed elsewhere \cite{Wr98,Wr92}.

The M9A beamline \cite{Al80} provided
a negative pion flux of typically $7 \times 10^{5}$~s$^{-1}$
with a central momentum $81.5$~MeV/$c$,
momentum bite 9\%,
and spot size $3.4$$\times$$4.4$~cm$^2$.
The beam had a 
microscopic time structure consisting 
of a 2-4~ns pulse width with a 43.3~ns pulse separation
and a 99\% macroscopic duty cycle.
For the hydrogen measurement we conducted the experiment
with the M9A radio frequency separator
which yielded a $\pi$/$e$/$\mu$ beam composition
of  0.73/0.18/0.09. 
For the deuterium measurement we conducted the experiment
without the separator
which yielded a $\pi$/$e$/$\mu$ beam composition
of 0.06/0.89/0.05.\footnote{This 
running condition resulted from the
failure of the RF separator in the D$_2$ experiment.}

The incident beam particles were counted 
in a 4-element plastic scintillator telescope
with a total thickness 0.635~cm
and stopped in either a liquid hydrogen
target or a liquid deuterium target.
Both targets had cylindrical cross sections
with dome-shaped front faces,
and were located at the geometrical center
of the detector system.
The H$_2$ cell was
15~cm in length, 16~cm in diameter, and 2.7~$\ell$ in volume, 
with Au front and side walls 
of thickness $0.25$~mm.
The D$_2$ cell was
15~cm in length, 15~cm in diameter, and 2.5~$\ell$ in volume,
with Cu front and side walls 
of thickness $0.50$~mm.
For the H$_2$ experiment the hydrogen filling 
had a deuterium contamination of order 10$^{-6}$ 
and for the D$_2$ experiment the deuterium filling
had a hydrogen contamination of order 10$^{-3}$.

The outgoing photons were detected by pair conversion
in a cylindrical Pb converter 
and $e^+$-e$^-$ tracking 
in two cylindrical tracking chambers.
The Pb converter had a thickness 1.12$\pm$0.01~mm, 
radius $13$~cm, and length $35$~cm,
and was centered
at the geometrical center
of the detector system.
The tracking chambers comprised
a low mass cylindrical multi-wire proportional chamber 
and a large volume cylindrical drift chamber.
The MWPC had an inner radius
26.5~cm, outer radius 27.5~cm, and active length $\sim$$75$~cm,
and consisted of 768 axial anode wires
between $2$$\times$$384$ helical inner and outer cathode
strips. It provided a single $(x,y,z)$ coordinate
for each track.
The drift chamber had an inner radius
$30.2$~cm, outer radius $58.0$~cm
and active length $\sim$$90$~cm,
and comprised four concentric layers 
of drift cells with six anode wires per drift cell.
Layers one, two and four 
were axial and provided $( x , y )$-coordinate tracking information.
Layer three
was stereo and provided the $z$-coordinate tracking information.
The entire detector system
was located in a large solenoidal magnet
that provided a 1.2~kG axial magnetic field.

Concentric rings of segmented plastic scintillators 
were employed for fast-triggering.
These scintillators comprised: 
the A/A$^{\prime}$-rings, located just inside the Pb converter radius, 
the C-ring, located just outside the Pb converter radius,
and the D-ring, located just outside the drift chamber radius.
The A/A$^{\prime}$-rings had four-fold segmentation
and inner radii of 12.2 and 12.6~cm respectively. 
They were used to veto any charged
particles that emerged from the vicinity of the target.
The C-ring had twelve-fold segmentation and an inner radius 15.8~cm.
The C scintillators were used to identify the $e^+$$e^-$ pairs 
from $\gamma$-ray conversion in the lead converter.
The D-ring had sixteen-fold segmentation and an inner radius $\sim$59.4~cm.
The D scintillators were used to identify the individual electron and positron 
tracks that emerged from the drift chamber. 

The drift chamber times were recorded by LeCroy 1879 
pipeline TDCs while the MWPC hits were recorded by
custom-built electronics \cite{Li78}.
The pulse height and timing information
from the beam counters and trigger scintillators
were recorded by CAMAC ADCs and TDCs.
The data acquisition was based on two SLAC scanner 
processors (SSPs) \cite{Br85} 
and a VAX 3000 workstation. One SSP handled the event readout
from the hardware to an internal memory, the other 
SSP handled the data transfer 
from the internal memory to the computer, while the computer handled 
the data storage, run control, online monitoring, {\it etc}.

\subsection{Trigger system}
\label{trigger}

The two-photon trigger was based upon 
the hit multiplicities and hit topologies  
in both the trigger scintillator rings 
and the drift chamber cells.
The trigger was designed for efficient identification of
photon pairs from double radiative capture 
and efficient rejection of 
single photons and coincident photons
from single radiative capture and pion charge exchange.
The trigger was implemented in four stages that involved:
trigger scintillator multiplicities (level one), 
trigger scintillator topologies (level two), drift chamber cell 
multiplicities (level three), and drift chamber cell 
topologies (level four).

Level one was based on the multiplicities 
of the hits in the trigger scintillators.
The level-one condition was
$\overline{A}$$\cdot$$\overline{A'}$$\cdot$$\geq$$2$$C$$\cdot$$\geq$$3$$D$, 
where $\overline{A}$$\cdot$$\overline{A'}$ imposed a veto on charged particles,
$\geq$$2$$C$ required a pair of $\gamma$-ray conversions,
and $\geq$$3$$D$ required a minimum of three tracks
to fully traverse the drift chamber.

Level two was based on the topologies of
the hits in the C-D trigger scintillators rings
and was implemented using a LeCroy 2372 memory
look-up unit (MLU) and a LeCroy 4508 programmable logic unit (PLU). 
First the MLU, using the recorded distribution of C-counters hits,
defined the allowed distribution of D-counter hits
that was consistent with the topology of a double radiative capture event.
Then the PLU, using the allowed distribution of D-counter hits,
determined the number of so-called ``valid Ds'', {\it i.e.} the
number of matches between allowed D-counters and fired D-counters.
A total of $\geq$~2 non-adjacent C-counter hits
were required to define an allowed distribution of D-counter hits,
and a total of $\geq$3~valid Ds were required to
fulfill the level two trigger requirement.

Note that two different versions 
of the level two trigger were implemented 
for the hydrogen experiment and the deuterium experiment.
In the hydrogen experiment, 
if two C-counters were back-to-back,
{\it i.e.} were located on the opposite sides
of the C-ring, no allowed distribution 
of D-counter hits was defined.
This permitted a reduction
in the trigger rate due to photon pairs 
from $\pi^o \rightarrow \gamma \gamma$ decay 
following $\pi^- p \rightarrow \pi^o n$ charge exchange.
In the deuterium experiment,
the branching ratio for charge exchange is 
small and therefore rejection
of back-to-back C-counters was unnecessary.

Level three was based on the multiplicities of the cell hits
in the drift chamber and was derived 
by custom-built OR boards \cite{Be90},
which indicated if a cell was hit,
and a custom-built analog hit counter \cite{Be91},
which accumulated the distribution of hits.
The trigger imposed minimum values on the number
of the individual cell hits in the second drift chamber layer (denoted $n_{2}$)
and the number of contiguous cell clusters
in the third and fourth  drift chamber layers (denoted $n_{3-4}$). 
The values of $n_{2}$ and $n_{3-4}$
were chosen for good efficiency in accepting
photon-pair events and good efficiency in rejecting
single photon events.
In the hydrogen measurement 
we collected some data (45\%)
with $( n_{2} ,  n_{3-4} ) \geq ( 3, 3 )$
and some data (55\%) with $( n_{2} ,  n_{3-4} ) \geq ( 4, 6 )$.
In the deuterium experiment
we collected all data with $( n_{2} ,  n_{3-4} ) \geq ( 4, 6 )$.

Level four was based on the topologies of the cell hits in
the first layer of the drift chamber.
This trigger was implemented in software
by the SSP  and was based 
on cell information
from the OR boards.
It rejected any event with back-to-back cell hits in  
drift chamber layer one
and was useful in the reduction of the event rate
from $\pi^o \rightarrow \gamma \gamma$ decay
following $\pi^- p \rightarrow \pi^o n$ charge exchange.
The level four trigger was enabled 
for the hydrogen measurement 
but disabled for the deuterium measurement

A dedicated $\pi^o$ trigger -- which allowed
the back-to-back topologies of $\gamma$-ray conversions 
from $\pi^o$ decays -- 
was used for measurements of the spectrometer acceptance,
resolution function, {\it etc.}
A dedicated beam trigger -- which generated events
from pre-scaled hits in beam counters --
was used for the determination of
the pion content of the incident beam.

\section{Data analysis.}
\label{analysis}

We report hydrogen and deuterium data
that were collected
in  two four-week running periods.
The hydrogen experiment recorded a 
total of $3.10 \times 10^{11}$ pion stops
and the deuterium experiment recorded a 
total of $3.07 \times 10^{11}$ pion stops.

Note that the running conditions for data taking in hydrogen 
and deuterium were different in several ways.
Firstly a separated beam was used in the
hydrogen experiment, while an unseparated beam 
was used  in the deuterium experiment 
(see Sec.\ \ref{setup}). 
For deuterium 
this resulted in a large electron beam rate
which therefore required greater attention
to counting the $\pi^-$ stops.
Secondly the MWPC was not operable 
at the large electron beam rate
of the deuterium experiment.
For deuterium
this resulted in less z-tracking information
and poorer z-tracking resolution.
Thirdly, the level two trigger and the level four trigger
were different in rejecting
back-to-back hit topologies for the hydrogen experiment
but accepting back-to-back hit topologies for the deuterium 
experiment. 

The data analysis involved several steps.
In step one we applied cuts to identify photon pairs 
that originated from the target.
In step two we applied cuts to select the 
photon pairs from double radiative capture
and reject the photon pairs from various background sources.
In step three we determined the contributions
of any residual background events
in the double radiative capture events.
Lastly, we determined the number of 
negative pion stops, and obtained the photon 
pair acceptance, in order to derive the
branching ratio for double radiative capture. 

\subsection{Identification of photon pairs}
\label{signal events}

The analysis procedure involved 
first identifying the various tracks in the tracking chambers,
then pairing the $e^+$$e^-$ tracks to make individual photons,
and finally pairing the individual photons to make photon pairs.
The criteria for identifying ``good tracks'' and making
``good photons'' are described below.

\subsubsection{Tracking cuts}

Each helical track is defined 
by a circle fit to the $(x,y)$ coordinates
and a straight-line fit to the $(\rho,z)$ coordinates
that were obtained from the chamber hits. 
The tracks fell into two categories:
so-called ``non-wrap around'' tracks 
where the particle emerges through
the drift chamber outer radius,
and so-called ``wrap around'' tracks 
where the particle spirals
within the drift chamber outer radius.
We imposed requirements on the
maximum variance ($S^2_{xy}, S^2_z$) 
and the minimum points ($N_{xy}, N_z$)
in the $( x , y )$ circle fit and $( \rho, z )$ straight-line fit.
In addition, for hydrogen,
a requirement was imposed 
on the maximum distance $d_{wc}$ 
of the particle's trajectory
from the associated MWPC hit.

The cut parameters for the hydrogen analysis
and the deuterium analysis are summarized in Table \ref{t: 1H tracking cuts}.
Note that the $( x , y )$ resolution
is somewhat poorer for positrons than electrons 
-- due to the 2.7$^{\circ}$ Lorentz angle of the drift electrons
in the 1.2~kG magnetic field --
and therefore the values of $S^2_{xy}$
are larger for $e^+$ tracks.
For hydrogen a cut was also imposed on the
distance $d_{wc}$ between the fitted track and the
corresponding MWPC hit.

\subsubsection{Photon cuts}

The $e^+$$e^-$ tracks from $\gamma \rightarrow e^+e^-$ conversion 
must converge at the Pb converter radius.
Therefore in pairing the $e^+$$e^-$ tracks into reconstructed photons
we imposed a maximum on the x,y separation ($d_{xy}$) 
and the z separation ($d_{z}$) of the $e^+$$e^-$ tracks at the Pb radius.
In addition, we required that the reconstructed 
photon must have originated from the direction of the target cell.
Therefore a cut was imposed on the magnitude ($r_{close}$) and the
z-component ($z_{close}$) of the closest approach
of the photon momentum vector to the detector's geometrical center

The values of cuts on $d_{xy}$, $d_{z}$, $r_{close}$ and $z_{close}$
for hydrogen and deuterium are summarized 
in Table \ref{t: 1H photon cuts}.
Note that the z-tracking was poorer
in the deuterium experiment
than the hydrogen experiment,
due to the inoperable MWPC.
This is reflected in the different values
for $d_{z}$ and $z_{close}$
in hydrogen and deuterium.

\subsubsection{Photon pair events}

Good photon pairs were defined as events with
four good tracks, {\it i.e.}\ four tracks 
passing all tracking cuts in Table \ref{t: 1H tracking cuts},
and two good photons, {\it i.e.}\ two photons 
passing all photon cuts in Table \ref{t: 1H photon cuts}.
From $3.10 \times 10^{11}$ pion stops
in hydrogen  a total of $2.3 \times 10^{6}$ 
photon pairs survived the combination of tracking cuts and photon cuts.
From $3.07 \times 10^{11}$ pion stops
in deuterium a total of $2.3 \times 10^{5}$ photon pairs 
survived the combination of tracking cuts and photon cuts.

\begin{figure}
\begin{center}
\epsfig{file=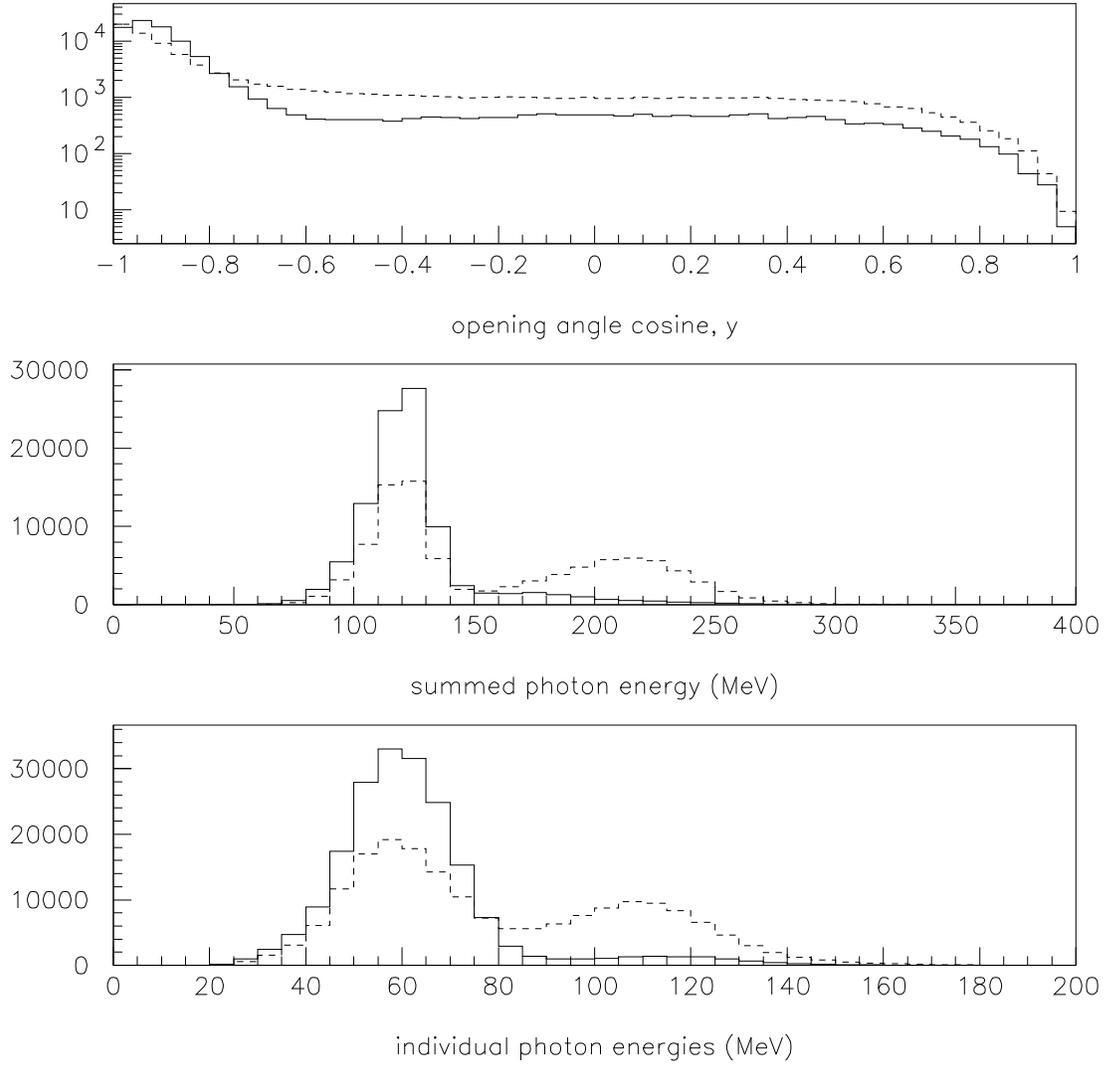,height=16.0cm}
\end{center}
\caption{The opening angle (top), summed energy (center) and 
individual energy (bottom) spectra 
of the two-photon events passing the tracking 
cuts and photon cuts in the hydrogen data-set (solid histogram)
and the deuterium data-set (dashed histogram). 
The $\pi^o$ background is apparent as the events with $y < -0.7$
in the opening angle spectrum and the $\pi$$\pi$ background 
is apparent as the events with $\omega_1 + \omega_2 > 150$~MeV in the 
sum energy spectrum. To facilitate their comparison
the hydrogen spectra and deuterium spectra are normalized to equal 
numbers of photon-pair events.} 
\label{fig:1h_2h_photonpairs}
\end{figure}

The summed energy ($\omega_1 + \omega_2$) spectra,
individual energy ($\omega$) spectra,
and opening angle cosine ($y$) spectra, 
for reconstructed photon pairs from hydrogen and deuterium, 
are shown in Fig.\ \ref{fig:1h_2h_photonpairs}.
The photon energies $\omega$ were obtained via the sum of the 
$e^+$$e^-$ momentum vectors at
the converter radius, {\it i.e.} 
$\vec{p}_{\gamma} = \vec{p}_{e^+} + \vec{p}_{e^-}$,
and the opening angle $y$ was obtained 
via the scalar product
of the two $\gamma$-ray momenta, {\it i.e.} 
$y = \vec{p}_{\gamma1} \cdot \vec{p}_{\gamma2} / 
|\vec{p}_{\gamma1}| |\vec{p}_{\gamma2}|$.
Note that the $e^+$,$e^-$ transverse momenta $p_t$ 
were derived from the track radius $r$ in the circle fit
via $p_t = e  r B$, and the $e^+$$e^-$ total momentum $p$ 
was derived using the track pitch $\phi$ from the 
straight-line fit via $p = p_t /\sin{\phi}$.

The spectra in Fig.\ \ref{fig:1h_2h_photonpairs} 
are dominated by real photon
coincidences from the $\pi^o$ background and
accidental photon coincidences from the $2$$\pi$
background. The $\pi^o$ background is most readily seen 
in the opening angle  spectra as events with 
$y < -0.7$ and the $2$$\pi$ background
is most readily seen in the sum energy spectra 
as events with $E > 150$ MeV. 
The different numbers of photon pairs 
from $^1$H and $^2$H -- {\it i.e.}\ 
$2.3 \times 10^{6}$ versus $2.3 \times 10^{5}$
-- reflects the much larger $\pi^o$ background 
in the $^1$H experiment than in the $^2$H experiment.

\subsection{Identification of double radiative capture events}

Next we describe the 
cuts we applied on the 
beam counter pulse height
and C-counter timing information
to remove the $2$$\pi$ and $\mu$$\pi$ backgrounds
and the opening angle 
to remove the $\pi^o$ background.
These cuts are denoted the
beam counter amplitude cut,
the C-counter timing cut
and the opening angle cut, respectively.

\subsubsection{Beam counter amplitude cut}

\begin{figure}
\begin{center}
\epsfig{file=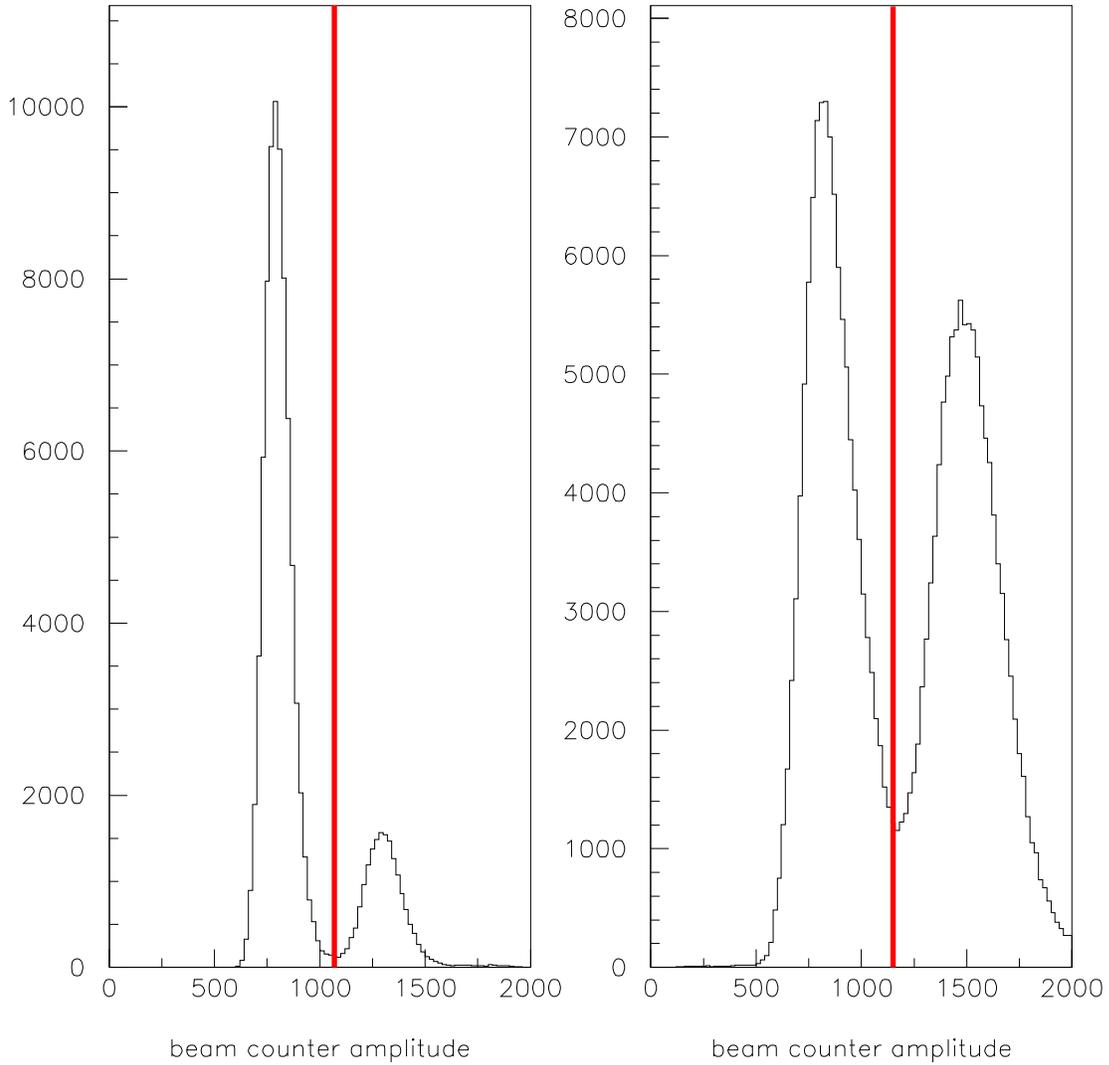,height=16.0cm}
\end{center}
\caption{The beam amplitude spectra for the
hydrogen data-set photon pairs (lefthand plot) and 
deuterium data-set photon pairs (righthand plot).
The plots show the single pion stops
associated with real photon pairs (the lower amplitude peak) 
and two pion stops associated with accidental photon pairs 
(the higher amplitude peak). The beam counter cuts, 
at Chan.\ 1069. in the hydrogen data-set and Chan.\ 1150 in
the deuterium data-set, are indicated by vertical bars.}
\label{fig:1h_2h_bmadc}
\end{figure}

To distinguish between single pion stops and 
multiple pion stops  we applied a cut on the
pulse height information from the four beam counters.
First, the pulse heights
from the eight photo-multipliers 
viewing the four beam scintillators 
were normalized and summed
to obtain a single beam counter amplitude.
The resulting amplitude spectra 
from representative $^1$H and $^2$H runs
are shown in Fig.\ \ref{fig:1h_2h_bmadc}.
The lower energy peak corresponds to events
with single $\pi^-$ stops
while the higher energy peak corresponds to 
events with two $\pi^-$ stops.
In the $^1$H spectrum 
the peak widths were dominated by the energy straggling 
of the incoming pions and the finite resolution of the beam counters.
In the $^2$H spectrum 
the peak widths were further increased by 
additional pile-up with beam electrons.

Also indicated in Fig.\ \ref{fig:1h_2h_bmadc} 
are the corresponding beam counter amplitude cuts
at Chan.\ 1069 in hydrogen and Chan.\ 1150 in deuterium.
The cut efficiencies for passing two-photon events 
from single $\pi^-$ stops were determined 
utilizing the photon pairs from
$\pi^o \rightarrow  \gamma \gamma$ decay
via both dedicated $\pi^o$ trigger runs
and standard two-photon trigger runs.
We found an efficiency of 100\% for
the hydrogen data and  92\% for the deuterium data.

\subsubsection{C counter timing cut}

Two categories of accidental $\gamma$$\gamma$ coincidences
are capable of circumventing the beam counter amplitude cut.
They are: (i) $\mu$$\pi$ background events
involving one prompt $\gamma$-ray
from a $\pi^-$ stop and one delayed $\gamma$-ray
from an earlier $\mu^-$ stop,
and (ii) $\pi$$\pi$ background events
originating from two separate $\pi^-$ stops 
in two adjacent beam pulses.
Note that the $\mu$$\pi$ background was
much larger in the unseparated-beam deuterium experiment 
than in the separated-beam hydrogen experiment.

\begin{figure}
\begin{center}
\epsfig{file=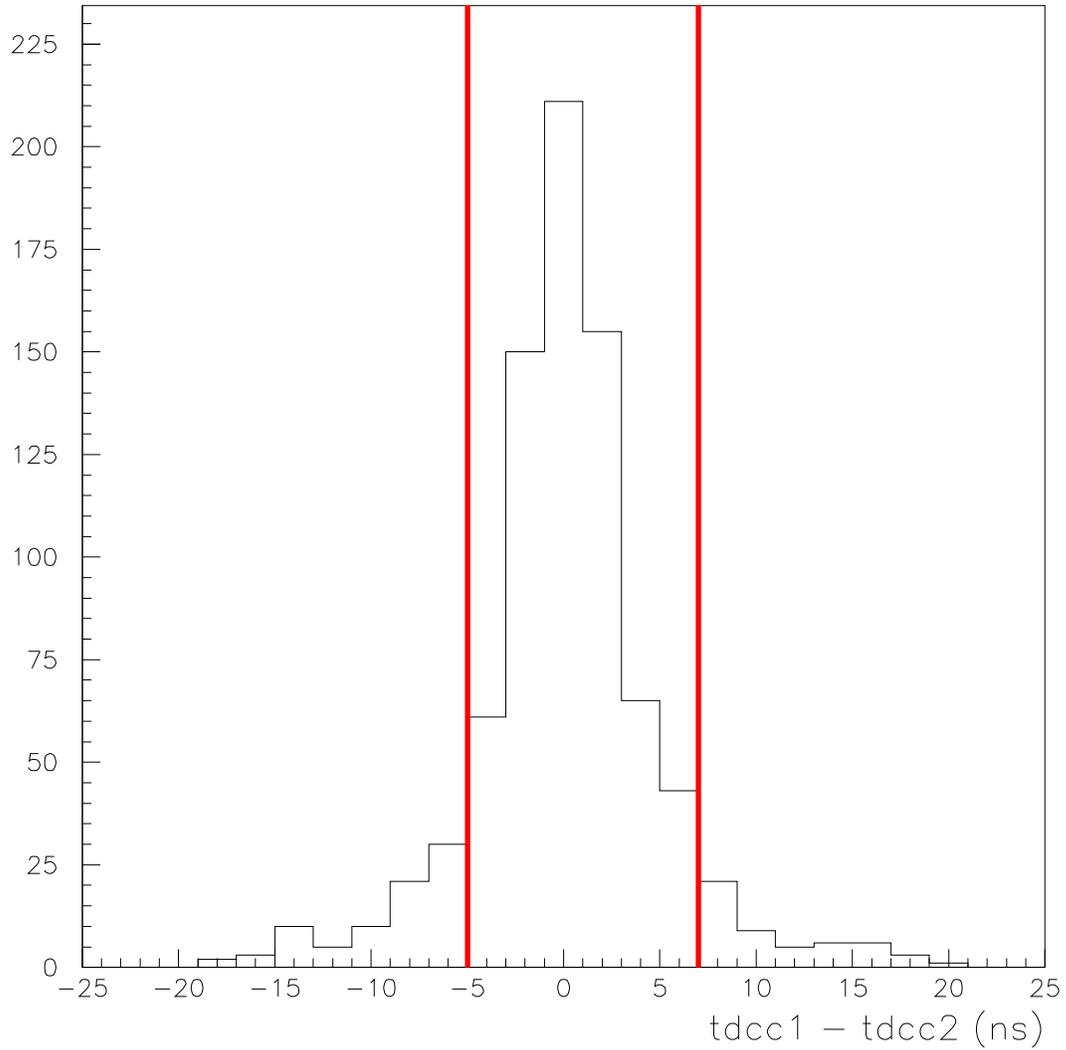,height=16.0cm}
\end{center}
\caption{The time difference (ns) between 
two C-counters hits for photon pairs in deuterium that pass
all cuts with the exception of the C-counter timing cut. 
The events outside the coincidence window $-5 < (t_{C1} - t_{C2} ) < +7$~ns
are dominated by $\mu$$\pi$ and $\pi$$\pi$ baclground.}
\label{fig:2h_tdcc1-tdcc2}
\end{figure}

To remove such backgrounds we applied a
C counter timing cut.
The cut was based on the time difference $(t_{C1} - t_{C2})$ 
between the two C-counters that were fired by the
two $e^+$$e^-$ pairs in the two-photon event.
For hydrogen we used a timing window of
$-4 < (t_{C1} - t_{C2}) < +4$~ns 
and for deuterium we used a timing window of
$-5 < (t_{C1} - t_{C2}) < +7$~ns
(the $t_{C1} - t_{C2}$ time difference spectrum for
deuterium photon pairs is shown 
in Fig.\ \ref{fig:2h_tdcc1-tdcc2}).
Their corresponding efficiencies for passing 
real $\gamma$$\gamma$ coincidences were determined 
utilizing the photon pairs from
$\pi^o \rightarrow  \gamma \gamma$ decay
using both dedicated $\pi^o$ trigger runs
and standard two-photon trigger runs.
We found an efficiency of 99.5\% for
the hydrogen data and 99.0\% for the deuterium data.

\subsubsection{Opening angle cut}

Finally, we applied a cut on the opening angle cosine $y$
to remove the $\pi^o$ background (the 
distributions in opening angle for all photon-pairs
from hydrogen and deuterium are shown in 
Fig.\ \ref{fig:1h_2h_photonpairs}).
In our final analyzes we used values 
of $y \geq -0.1$ for hydrogen and deuterium
although studies were made of the sensitivity
of our results to the cut value.
Note that (i) the $\pi^o$ background was
much larger in the hydrogen measurement and (ii) 
the opening angle resolution was significantly poorer 
in the deuterium measurement. 

The efficiency of the opening cut for passing events
from double radiative capture is obviously
dependent on the angular distribution
for the $( \pi , \gamma \gamma )$ events. 
The cut's effect on  $( \pi , \gamma \gamma )$ events
is discussed in detail in Sec.\ \ref{results}.

\subsubsection{Double radiative capture events}

The resulting distributions of events
in individual energy, summed energy and opening angle 
-- after the tracking, photon, beam counter amplitude, C
counter timing and opening angle cuts --
are shown in Fig.\ \ref{fig:1h_2h_signal} for hydrogen
and deuterium.
For hydrogen the spectra contain a total 
of 665 two-photon events 
with 597 events having sum energies $<$$150$~MeV
and 566 events having sum energies $80$-$150$~MeV
(the sum energy cut $<$$150$~MeV removes 
the vast majority of the $\pi$$\pi$ background events
and the sum energy cut $>$$80$~MeV removes
the vast majority of any Dalitz background events).
For deuterium the spectra contain a total 
of 521 two-photon events 
with 335 events having sum energies $<$$150$~MeV
and 327 events having sum energies $80$-$150$~MeV.
 
The summed photon energy spectrum shows a peak
at $E \sim m_{\pi}$ that corresponds to 
the double radiative capture events.\footnote{The peak centroid is
shifted downwards due to energy loss of the $e^+$$e^-$ pairs
in traversing the lead converter, trigger scintillators, etc.} 
The photon pairs with summed energies $> 150$~MeV
are dominated by accidental $\gamma$$\gamma$ coincidences from
unrejected $\pi$$\pi$ background.
Note that the energy distribution of the 
accidental $\gamma$$\gamma$ coincidences
are different in hydrogen and deuterium -- the former being dominated
by gamma-rays following $( \pi^- , \pi^o )$ charge exchange,
the latter being dominated by gamma-rays 
following $( \pi^- , \gamma )$ radiative capture.
The individual photon energy spectrum 
shows a broad continuum
with a low energy cut-off at about 20~MeV 
from the acceptance of the spectrometer.
At individual photon energies $> 100$~MeV
the spectra are also dominated by the
accidental $\gamma$$\gamma$ coincidences 
from the  $\pi$$\pi$ background.

\begin{figure}
\begin{center}
\epsfig{file=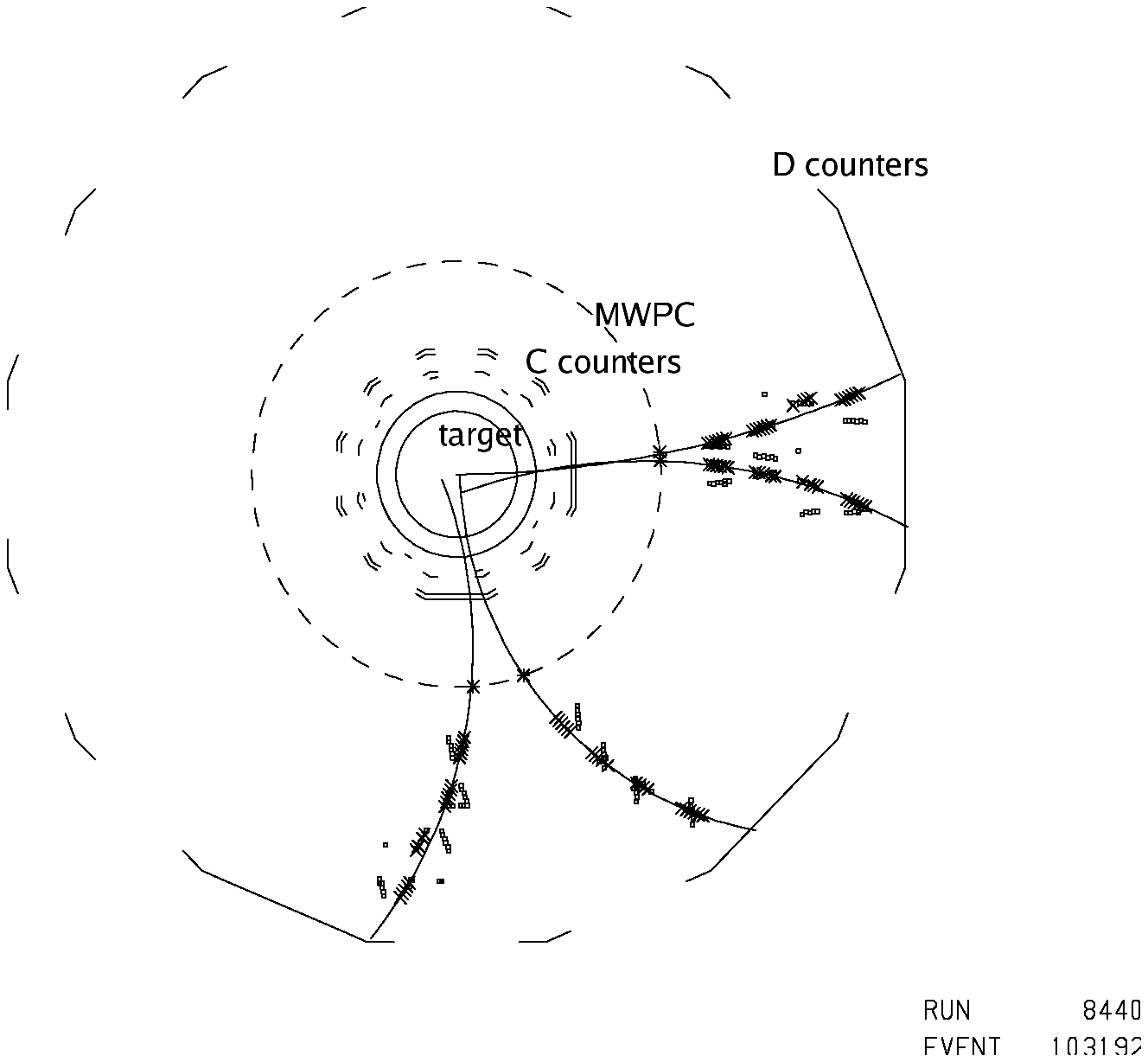,height=16.0cm}
\end{center}
\caption{A typical double radiative capture event
showing the trajectories of the two $e^+$$e^-$ pairs,
the circle fits to the four $e^+$$e^-$ tracks,
and the fired C and D counters.}
\label{fig:event}
\end{figure}

A typical double radiative capture event
is shown in Fig.\ \ref{fig:event}.
It has zero hits in the A-counter ring,
two hits in the C-counter ring,
and four hits in the D-counter ring.
The topology of trigger scintillator hits and 
drift cell hits is consistent 
with diverging $e^+$$e^-$ pairs,
and therefore the event fulfilled the
two-photon trigger.

\subsection{Subtraction of background events}
\label{background events} 

A small fraction of $\pi^o$ background events 
and $\pi$$\pi$ background events
were found to survive the beam amplitude, 
C-counter timing and opening angle cuts 
-- and therefore to contaminate the sum energy region $<$$150$~MeV of
double radiative capture events. Below we describe the methods 
we used to subtract these backgrounds.

\subsubsection{$\pi^o$ background}
\label{pi0 background subtraction}

To estimate the number of $\pi^o$ background events 
that survived the cuts
we utilized (i) the measured opening angle distribution
for $\pi^o$ events at angles $y < -0.45$ 
and (ii) the simulated opening angle spectrum
for $\pi^o$ events at all angles. 

The Monte Carlo simulation was performed using
the CERN GEANT3 package \cite{GEANT} (see Sec.\ \ref{acceptance} for details).
Recall, the true angular distributions of 
$\gamma$-$\gamma$ coincidences from $\pi^o$ decay
following at-rest charge exchange on hydrogen
are limited to $-1.0 < y < -0.91$ (the $\pi^o$'s originating
from charge exchange on hydrogen dominate 
both the hydrogen data-set and the deuterium data-set).
However, the measured angular distribution
of $\gamma$-$\gamma$ coincidences from $\pi^o$ decay
shows a significant opening angle tail
from the pair spectrometer response function.

\begin{figure}
\begin{center}
\epsfig{file=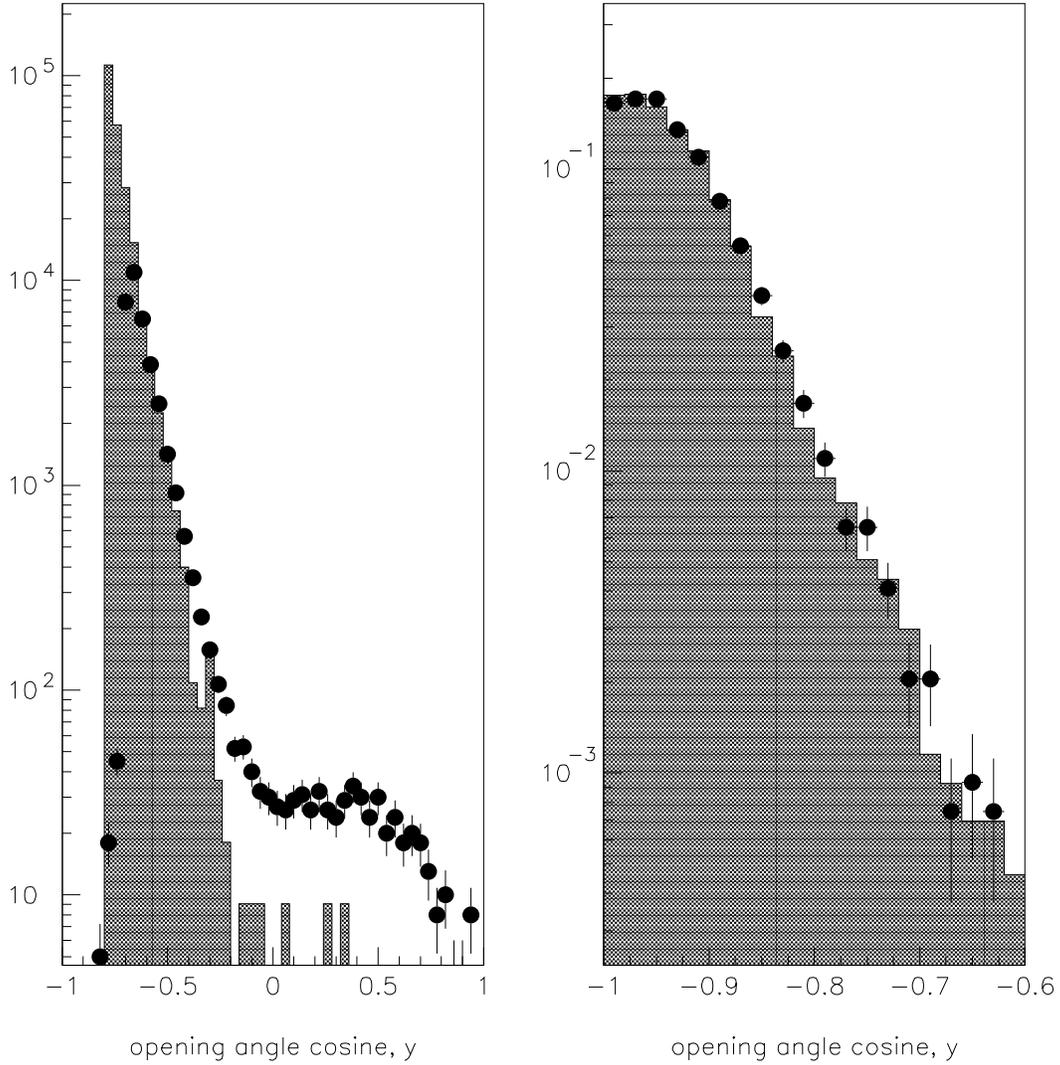,height=16.0cm}
\end{center}
\caption{Comparison of Monte Carlo and experimental
data for photon-pairs from $\pi^o$ decay for the
$\gamma$$\gamma$ trigger and $-0.8 < y +1.0$ (lefthand plot)
and the $\pi^o$ trigger and $-1.0 < y -0.6$ (righthand plot).
The solid histograms are the Monte Carlo results and the 
data points are the experimental results. (The apparent difference 
between Monte Carlo and experimental data for $y < -0.75$ 
in the $\gamma$$\gamma$ trigger data is a consequence of a 
loose y-cut that was applied in the initial processing of
the experimental data )}
\label{fig:pi0tail}
\end{figure}

Fig.\ \ref{fig:pi0tail} compares
the measured and simulated opening angle spectra
for the region $-1.0 < y < -0.6$ 
using a dedicated $\pi^o$ trigger (Sec.\ \ref{trigger}).
It indicates good agreement
between measurement and simulation
for $\pi^o$ events,
and affirms our confidence 
in using the simulation 
to determine the angular response of the pair spectrometer
and to estimate the $\pi^o$ contamination in the region $y>-0.1$.

Fig.\ \ref{fig:pi0tail} also compares the angular distribution
from measurement and simulation
for the region $-0.8 < y < +1.0$ 
using the standard 2$\gamma$ trigger (Sec.\ \ref{trigger})
for the photon-pairs surviving all the applied cuts except the angle cut.
The simulated data in Fig.\ \ref{fig:pi0tail} imply 
a small fraction of $\pi^o$ events
with reconstructed opening angles
$y > -0.1$ -- thus contaminating the region
of double radiative capture events.
To estimate the $\pi^o$ contamination
we defined two windows:
a $\pi^o$ window of $-0.7 < y < -0.45$ 
in which the $\pi^o$ background is dominant,
and a $(\pi , 2\gamma)$ window of 
$-0.1 < y < +1.0$ in which the $(\pi , 2\gamma)$ signal is dominant.
Using these windows the $\pi^o$ contamination with $y > -0.1$
was determined via

\begin{equation}
N^{EX}_{2\gamma} = N^{EX}_{\pi^o} ( N^{MC}_{2\gamma} / N^{MC}_{\pi^o} )
\end{equation}

where $N^{EX}_{\pi^o}$
is the number of measured events in the $\pi^o$ window
and $N^{MC}_{2\gamma}$ and $N^{MC}_{\pi^o}$
are the number of simulated events in the 
$\pi^o$ and $(\pi , 2\gamma)$ windows.
For hydrogen
the procedure yielded a $\pi^o$ background estimate
of $53 \pm 30$ events.
and for deuterium
the procedure yielded a $\pi^o$ background estimate
of $2 \pm 1$ events.
Note that the errors in the estimate of the $\pi^o$ background
are dominated  by the statistical uncertainties 
in the $\pi^o$ simulation.
The simulation was based on
generating a total of 9$\times$10$^7$
$\pi^o$ decays from at-rest charge exchange.

\subsubsection{$\pi$$\pi$ background}
\label{random background subtraction}

\begin{figure}
\begin{center}
\epsfig{file=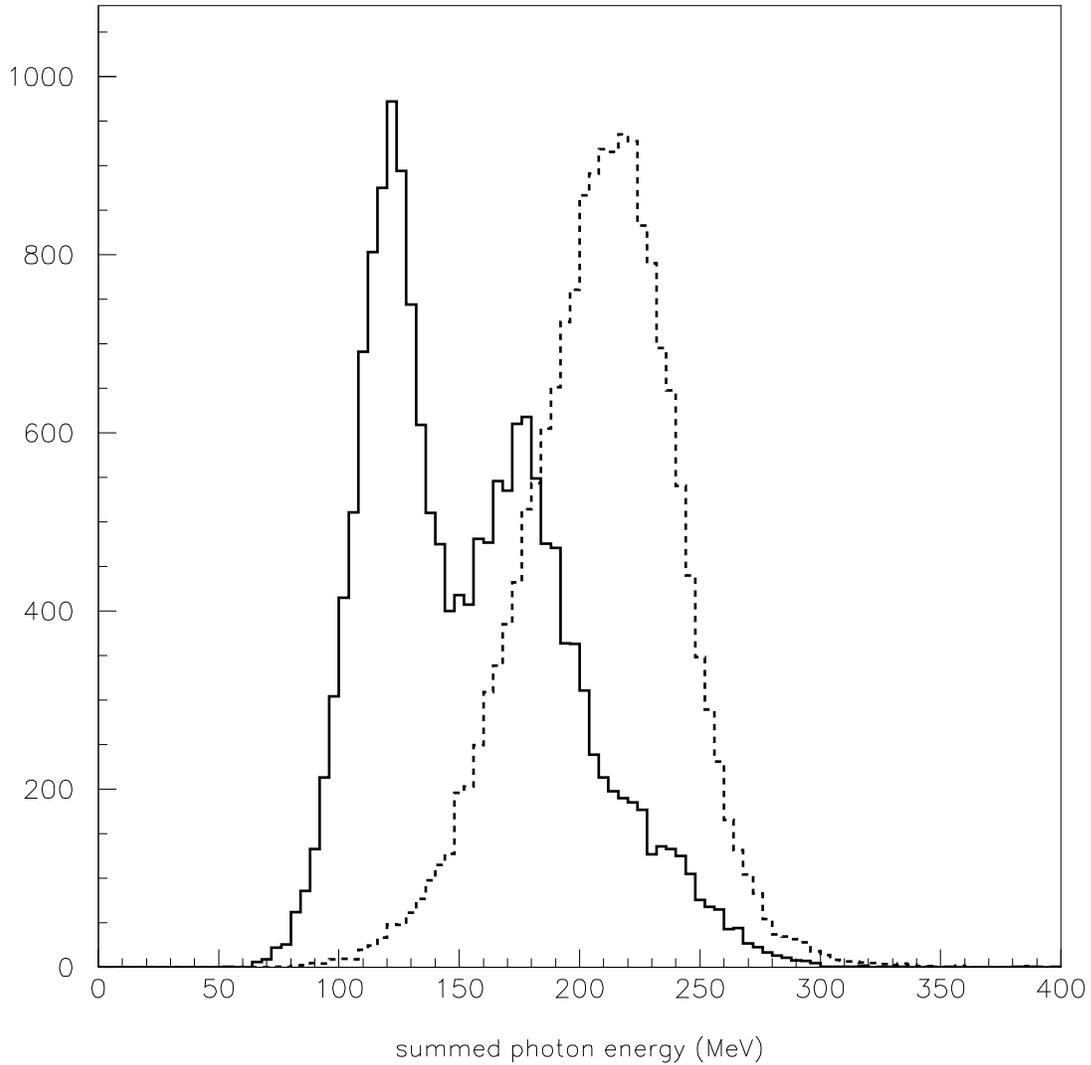,height=16.0cm}
\end{center}
\caption{The photon sum energy spectrum
for hydrogen (solid curve) and deuterium (dashed curve)
for events that fail the beam amplitude cut,
{\it i.e.} accidental $\gamma$$\gamma$ coincidences 
due to multiple pions in a single beam pulse.
In deuterium the accidental coincidences are dominated 
by single radiative capture.
In hydrogen the accidental coincidences originate 
from pion charge exchange
and single radiative capture.}
\label{fig:1h_2h_2pi}
\end{figure}

To estimate the number of $\pi\pi$ background events 
that survived the cuts we utilized their measured sum energy spectra 
in Fig.\ \ref{fig:1h_2h_2pi}. 
The summed energy spectra
were obtained by selecting 
the photon pairs failing 
the beam counter amplitude cut. 
In the hydrogen spectrum (Fig.\ \ref{fig:1h_2h_2pi} solid curve), 
the lower energy peak at $\sim$120~MeV 
is predominantly two $\gamma$-rays from 
two $\pi^o \rightarrow \gamma \gamma$ decays
while the higher energy peak at $\sim$180~MeV 
is predominantly one $\gamma$-ray from $\pi^o \rightarrow \gamma \gamma$ decay
and another $\gamma$-ray from single radiative capture.
In the deuterium spectrum (Fig.\ \ref{fig:1h_2h_2pi} dashed curve),
the single broad peak at $\sim$230~MeV originates from
two $\gamma$-rays from two single radiative capture events.

\begin{figure}
\begin{center}
\epsfig{file=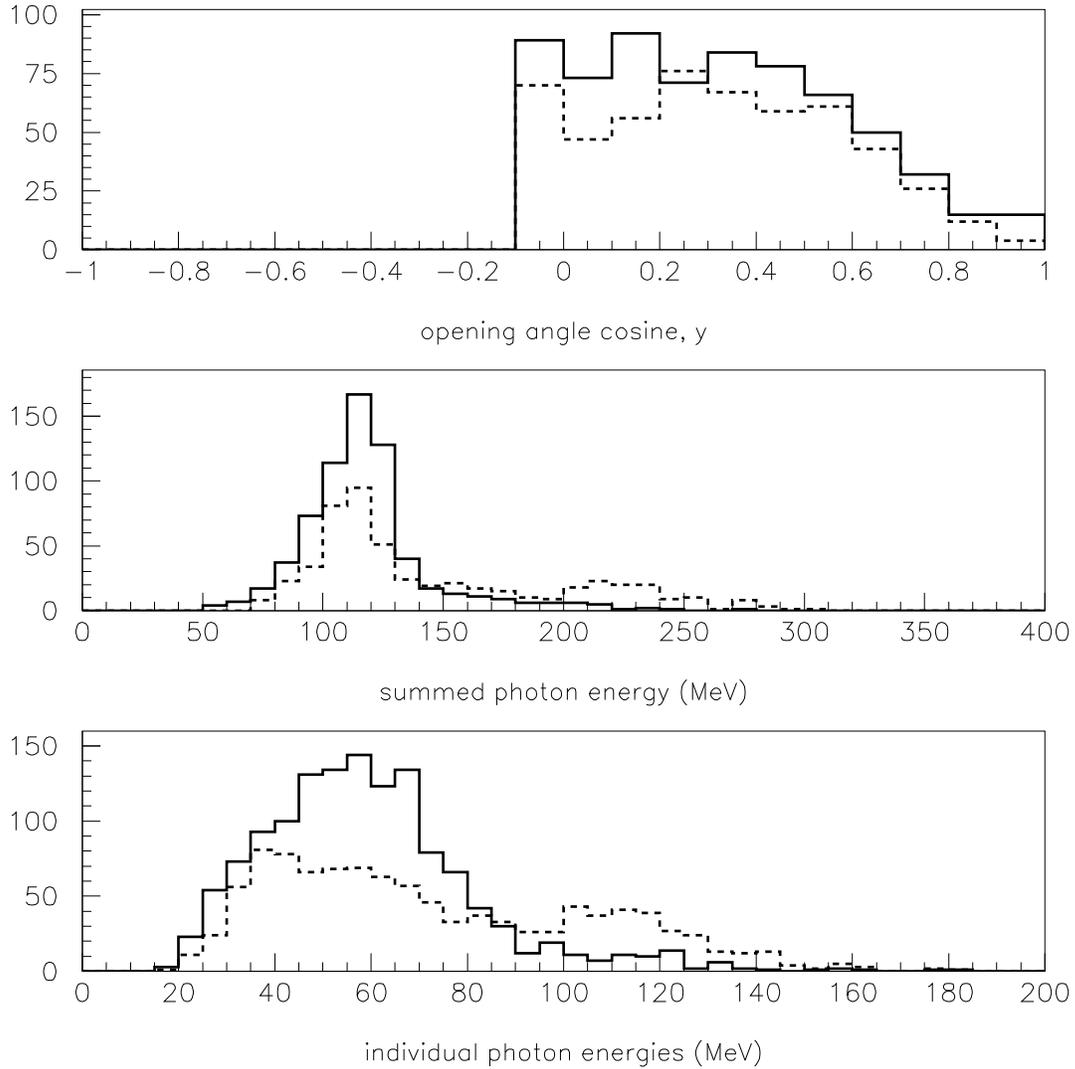,height=16.0cm}
\end{center}
\caption{The opening angle (top), summed energy (center) and individual
energy (bottom) spectra for photon pairs after all cuts
for the hydrogen data-set (solid histogram) 
and deuterium data-set (dashed histogram). 
The hydrogen spectra contain a total
of 665 photon-pair events with $y > -0.1$ 
The deuterium spectra contain a total
of 521 photon-pair events with $y > -0.1$ }
\label{fig:1h_2h_signal}
\end{figure}

Clearly the sum energy of $\pi\pi$ events 
can exceed the kinematical limit for
double radiative capture events.
We therefore defined an upper threshold $E_{\circ}$,
above which double radiative capture events are absent 
but $\pi\pi$ background events are present,
and computed the number of $\pi\pi$ events 
in the double radiative capture region $<$150~MeV
from number of $\pi\pi$ events 
in the high energy region $>$$E_{\circ}$.
For hydrogen this method gave
48$\pm$8 $\pi$$\pi$ background events below 150~MeV
in the double radiative capture sum energy spectrum
of Fig.\ \ref{fig:1h_2h_signal}.
For deuterium the same method gave
20$\pm$2 $\pi$$\pi$ background events below 150~MeV
in the double radiative capture sum energy spectrum
of Fig.\ \ref{fig:1h_2h_signal}.
The total number of $\pi$$\pi$ events at all energies
were $\sim$100 in hydrogen and $\sim$200 in deuterium.

A slight complication
in this approach
was the finite resolution 
of the pair spectrometer.
This generates a small high energy tail 
of double radiative capture events 
that extends beyond their kinematic limit. 
We therefore varied the upper threshold $E_{\circ}$ 
from $160$ to $180$~MeV to examine
the sensitivity to effects of the high energy tail
of the spectrometer resolution function.
For both hydrogen and deuterium the resulting variations
in the resulting estimates of the $\pi$$\pi$ contamination
were smaller than 1$\sigma$.

\subsubsection{Other backgrounds}

One potential source of background events 
was accidental coincidences between a
delayed photon from $\mu$ decay
and a prompt photon from $\pi$ capture
-- the $\mu$$\pi$ background. 
Such events 
were present
in the unseparated-beam deuterium measurement
but absent
in the separated-beam hydrogen measurement.
In deuterium, the contamination
from $\mu$$\pi$ events that survived
the C-counter timing cut
was estimated   
from the $\mu$$\pi$ continuum background
in the $t_{C1}$-$t_{C2}$ time spectrum.
This procedure yielded a
background estimate
of $( 15 \pm 4 )$ $\mu$$\pi$ events
in the deuterium data set.

Another potential source of background events
was $\pi^o \rightarrow \gamma e^+e^-$ Dalitz decay 
following $\pi^- p \rightarrow \pi^o n$ charge exchange
with the  combination
of the Dalitz pair
and the $\gamma$ conversion
yielding a photon-pair event.
Although the branching ratio 
for the $\pi^o \rightarrow \gamma e^+e^-$ decay mode
is small (1.2\%) and the inefficiency 
of the $A$,$A^{\prime}$ veto
is tiny ($<$1$\times$10$^{-4}$),
it is possible for Dalitz events
to generate photon pairs with small opening angles.
We therefore conducted a GEANT3 simulation
of Dalitz events which showed that
the Dalitz events with small opening angles ($y > -0.1$)
had small summed energies ($<$80~MeV).
Consequently, a worst-case estimate of the 
Dalitz background in the sum energy range $80$-$150$~MeV
was made by attributing all events with $<$80~MeV,
to Dalitz background, giving a worst-case contribution of 
$<2$ events in hydrogen and $<1$ event in deuterium. 

Another potential source of background events 
is nuclear double radiative capture
from pion stops in neighboring materials. 
Using (i) the fraction $(15 \pm 1)$\% of
pion stops in materials other than liquid H$_2$/D$_2$ (see Sec.\ \ref{results}),
and (ii) an estimated branching ratio $5 \times 10^{-6}$ 
for  nuclear double radiative capture with E$>25$~MeV 
and $y > -0.1$ \cite{De79,Ma80},
we derived upper limits
of $<$10  nuclear $(\pi , 2 \gamma )$ events
in the hydrogen data-set and
$<$20  nuclear $(\pi , 2 \gamma )$ events
for the deuterium data-set.
In addition, the fiducial cuts on photon trajectories
will likely remove a large fraction of such events
from pion stops in the  target walls, beam scintillators, 
trigger scintillators, {\it etc}.
We therefore treated the above estimates
as very conservative upper limits
on the nuclear $( \pi , 2 \gamma )$ background.

\subsubsection{Background-subtracted events.}
\label{background subtracted events}

A compilation of the various estimates of
the background sources in the hydrogen experiment
and the deuterium experiment are given in Table \ref{t: counts}.
After subtracting the backgrounds, we obtained 
a total of 465$\pm$$^{39}_{40}$ double radiative capture events 
with summed energies from 80 to 150~MeV in hydrogen 
and a total of 292$\pm$$^{19}_{27}$ double radiative capture events 
with summed energies from 80 to 150~MeV in deuterium.\footnote{Note
that in Ref.\ \cite{Tr02} we quote 482$\pm$42 as the number
of double radiative capture events with summed energies $>$80~MeV rather 
than 465$\pm$$^{39}_{40}$ events with summed energies 80$-$150~MeV.
However, the data set is identical.} 
The quoted uncertainties include the statistical
errors and the systematic errors in the signal counts
and the background estimates. 

\subsection{Counting of incident pions}
\label{pion counting}

The determination of the total number of incident pions 
was made by counting
the total number of incident beam particles 
and measuring the pion fraction of the incident beam.
The latter measurement was performed utilizing
the beam trigger (Sec.\ \ref{trigger})
and resulting amplitude and timing information 
from the beam counters. Note that the beam trigger events 
were collected simultaneously with the 2$\gamma$ trigger
events for the duration of the measurements.

\begin{figure}
\begin{center}
\epsfig{file=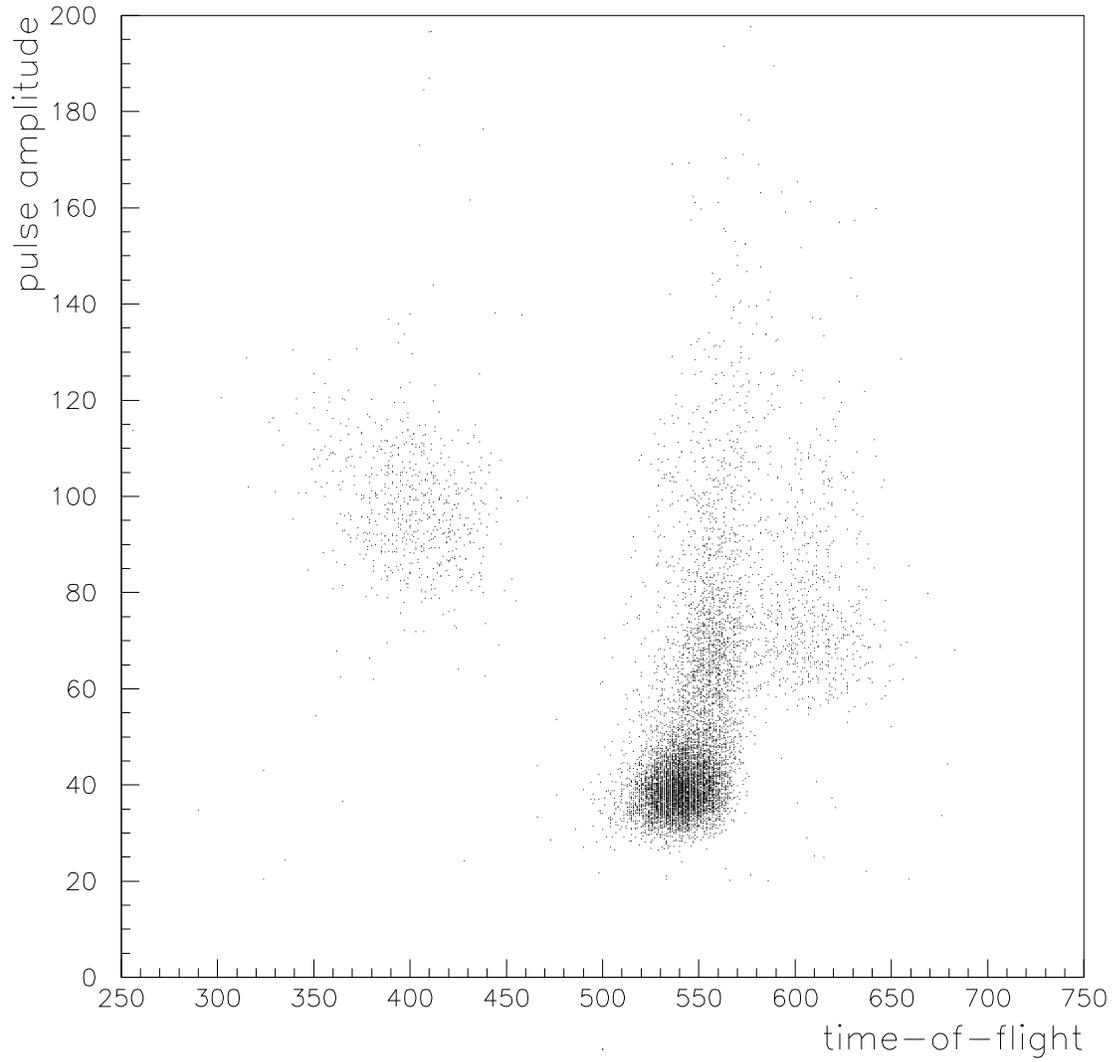,height=16.0cm}
\end{center}
\caption{Beam counter amplitude versus time-of-flight for
incoming beam particles in the deuterium experiment.
The pions are top-left, electrons are bottom-center
and muons are right-center.}
\label{fig:tofvamp}
\end{figure}

Because of the 43~ns micro-time structure 
of the primary proton beams
the arrival of secondary pions, muons and electrons 
are separated in time. 
At 81.5~MeV/c the FWHM of
pion arrival times 
was roughly 4.6~ns, while the separation 
from electron and muon arrivals was $\sim$14 
and $\sim$21~ns, respectively. 
In addition, 
the different particles 
have different energy loss 
and therefore different pulse heights
in the beam counters.
The combination of arrival time and 
pulse height at the beam counters was sufficient to 
clearly differentiate between incoming particles
in both the separated beam and the unseparated beam.
A representative plot of arrival time versus pulse height
for deuterium -- showing the quality of 
separation between pions, muons and electrons
-- is shown in Fig.\ \ref{fig:tofvamp}.

Based on the pulse height and time discrimination
between particle types we determined
an average beam composition for the hydrogen experiment
of 73\% $\pi^-$, 18\% $e^-$ and 9\% $\mu^-$
and the deuterium experiment of
6\% $\pi^-$, 89\% $e^-$ and 5\% $\mu^-$.
These pion fractions were used to derive
the number of livetime-corrected incident pions in the
hydrogen experiment of $( 3.10 \pm 0.03 ) \times 10^{11}$ and 
the deuterium experiment of $( 3.07 \pm 0.15 ) \times 10^{11}$.
The quoted uncertainties are based on the
analysis of the pion fraction in the incident beam
and are larger for the unseparated-beam deuterium 
experiment than for the separated-beam hydrogen experiment.

\subsection{Acceptance of pair spectrometer}
\label{acceptance}

To compare the experimental data
to theoretical calculations one requires
a determination
of the two-photon response
of the detector system.
The response function represents the probability 
of reconstructing photon pairs 
with true energies $\omega_1$, $\omega_2$ and opening angle $y$ 
with measured energies $\omega_1^{\prime}$, $\omega_2^{\prime}$
and opening angle $y^{\prime}$,
thus yielding the measured distribution of photon pairs
from the true distribution of photon pairs.
The response includes
the detector geometry, the interaction and conversion of photons, 
the energy loss and multiple scattering of electrons,
and the effects of the hardware trigger and the software
cuts. 

To determine the detector response function 
we employed a Monte Carlo simulation (RMCGEANT)
that was based on the CERN GEANT3 package \cite{GEANT}.
It incorporated the full geometry
of the detector system -- {\it i.e.}\ beam counters, 
target assembly,  trigger counters, tracking chambers
and the surrounding magnet --
and all relevant interactions
of photons and electrons.
The simulation produced data files 
containing Monte Carlo events 
that were analyzed 
with the same software
and the same cuts 
as the measured data. 
The program included event generators
for pion charge exchange $\pi^- p \rightarrow \pi^o n$, 
single radiative capture $\pi^- p \rightarrow \gamma n$
and double radiative capture $\pi^- p \rightarrow \gamma \gamma n$.
The generator we used for double radiative capture
was based on formulas contained in Beder's article \cite{Be79}.

An important test 
of the GEANT simulation 
of the detector system was afforded 
by the photon pairs 
from the $\pi$$\pi$ background.
These pairs originate from separate 
$\pi^- p \rightarrow \gamma n$ and/or 
$\pi^- p \rightarrow \pi^0 n$  reactions
from two pion stops in one beam pulse.
Since the absolute yields and energy-angle distributions
of such pairs are well known, 
they offer an invaluable means 
of calibration across 
the full range of photon energies and opening angles.
In the hydrogen experiment this calibration procedure
was performed continuously
for the entire duration of the data taking.
In the deuterium experiment the procedure
was limited to dedicated calibration runs 
with liquid H$_2$ filling.\footnote{During
running on deuterium the random $\gamma$-$\gamma$ coincidences
from the $\pi$$\pi$ background and true $\gamma$-$\gamma$ coincidences
from the $\pi^o$ background were used to monitor the 
relative acceptance for photon pairs. An absolute determination
of the acceptance was not possible due
to uncertainties in the hydrogen contamination 
in the liquid deuterium.}

\begin{figure}
\begin{center}
\epsfig{file=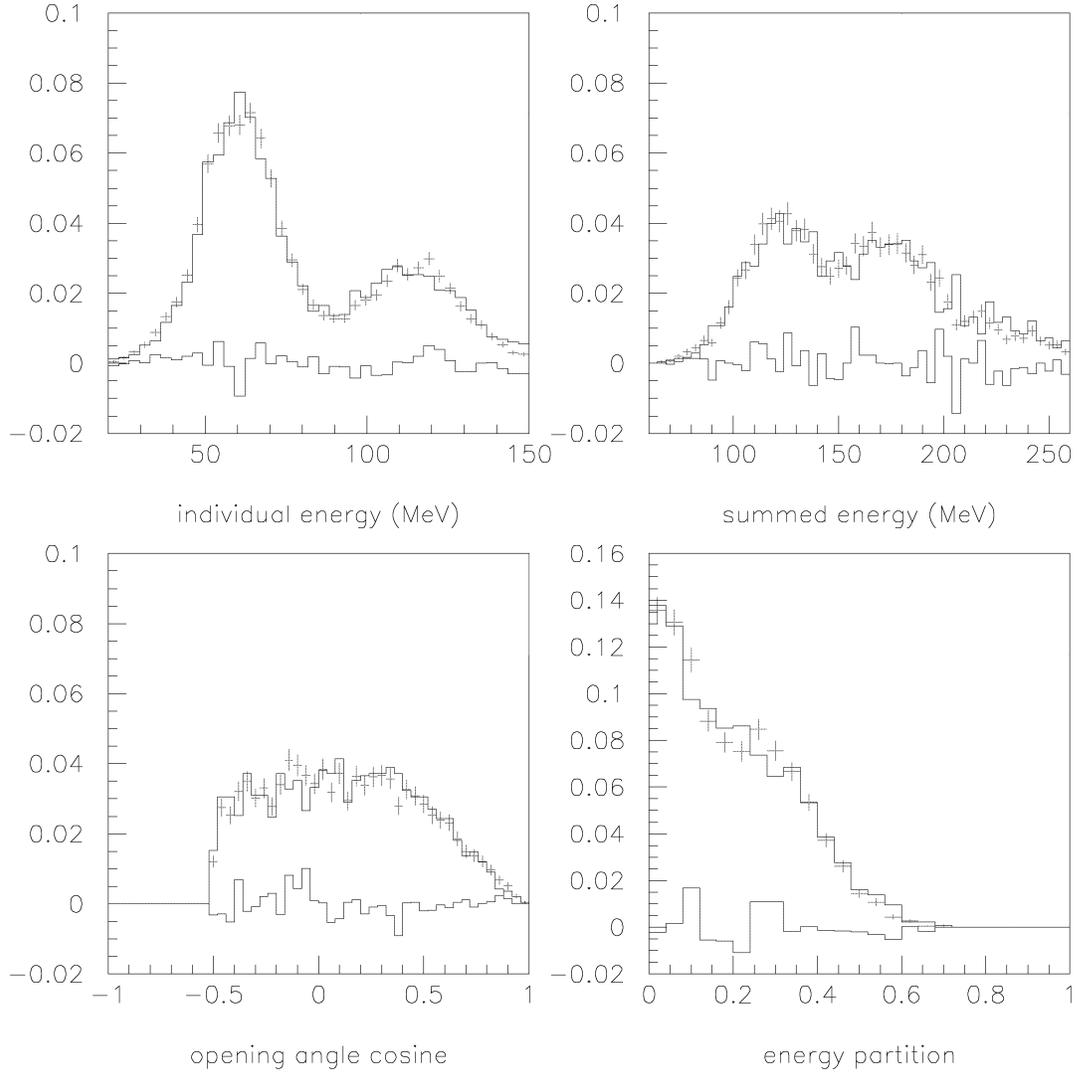,height=16.0cm}
\end{center}
\caption{Comparison of simulated data (solid line) 
and measured data (data points)
for $\pi$$\pi$ background events.
The plots are the individual photon energies (upper left),
summed photon energies (upper right), opening angle (lower left)
and energy partition (lower right).
Also show is the difference between the simulated data and the
measured data.}
\label{fig:expt_simu_rand}
\end{figure}

Representative measured and simulated spectra for
photon-pairs from $\pi$$\pi$ events
are compared in Fig.\ \ref{fig:expt_simu_rand}.
The energy-angle distributions from experiment and simulation
are in agreement at the level of a few percent
and affirms the high quality 
of the GEANT simulation
of the detector system. 
Note that the absolute efficiency 
for detecting photon pairs 
was found to be smaller in the experiment than the
simulation by a factor $F = 0.90 \pm 0.09$ in hydrogen 
and $F = 0.96 \pm 0.06$ in deuterium.
This difference is believed to originate
from detector inefficiencies that are present 
in the experiment but are absent from the simulation. 
These inefficiencies include 
trigger scintillator inefficiencies,
tracking chamber inefficiencies, 
and possible effects from chamber noise
on track recognition.
For both hydrogen and deuterium the run-by-run variation of the normalization 
factor was $\pm$4\% and the total change 
over the running period was $<$10\%.

Another important test of the overall quality of the GEANT simulation 
of the detector system was afforded by photon pairs from
$\pi^o \rightarrow \gamma \gamma$ decay
following $\pi^- p \rightarrow \pi^o n$ charge exchange.
These data were collected in runs using
a dedicated $\pi^o$ trigger in both the hydrogen experiment
and the deuterium experiment.
Likewise the measured data and simulated
data for photon pairs from $\pi^o \rightarrow \gamma \gamma$ decay
show good agreement 
and this also affirms
the high quality
of the GEANT simulation
of the detection system.
The normalization factors of $F = 0.90$
in hydrogen and $F = 0.96$ in deuterium
that were derived from the $\pi$$\pi$ study
are also consistent with the $\pi^o$ study.

The detector acceptance $\epsilon\Delta\Omega$
for double radiative capture 
was obtained using the RMCGEANT package 
and the double radiative capture event generator.
The procedure yielded acceptances of 
$\epsilon\Omega = 0.66 \times 10^{-4}$
for hydrogen and $\epsilon\Omega = 0.90 \times 10^{-4}$
for deuterium when using the standard energy-angle cuts
and assuming photon-pairs with energy-angle
distributions derived by Beder \cite{Be79}.
Note that the acceptance for deuterium is larger
than hydrogen due to the looser cuts
that were applied in the presence of the smaller $\pi^o$ background.

\section{Results}
\label{results}

Herein we discuss the experimental results from the
hydrogen experiment and the deuterium experiment.
We first describe the determination of the 
branching ratios in the two experiments
and then discuss the comparison between theory
and experiment for the elementary process
and the nuclear process.

\subsection{Branching ratios for hydrogen and deuterium}
\label{br}

The absolute branching ratios for double radiative capture 
on pionic hydrogen and pionic deuterium were obtained via
\begin{equation}
\label{e:br}
B.R. = \frac{N_{\gamma\gamma}}
{ N_{\pi^-} \cdot \epsilon\Delta\Omega \cdot F
\cdot c_{stop} \cdot c_{bm}} 
\end{equation}
where $N_{\gamma \gamma}$ is the number 
of background-subtracted double radiative capture events,
$N_{\pi^-}$ is the number of livetime-corrected incident pions,
$\epsilon\Delta\Omega \cdot F$ is the detector acceptance,
and $c_{stop}$ and $c_{bm}$ are minor correction factors
which are discussed below.

The values employed in computing the branching ratios
for hydrogen and deuterium are summarized in Table \ref{t: 1H 2pi bkd}.
For the double radiative capture events
with sum energies $80 < \omega_1 + \omega_2 < 150$~MeV 
and opening angles $y > -0.1$  
we took $N_{\gamma \gamma} = 465$$\pm^{39}_{40}$ for hydrogen 
and $N_{\gamma \gamma} = 292$$\pm^{19}_{27}$ for deuterium
from Sec.\ \ref{background subtracted events}.
For the livetime-corrected pion stops
we took  
$N_{\pi} = ( 3.10 \pm 0.03 ) \times 10^{11}$ for hydrogen 
and $N_{\pi} = ( 3.07 \pm 0.15 ) \times 10^{11}$ for deuterium
from Sec.\ \ref{pion counting}.
The GEANT simulation of detector acceptances
of Sec.\ \ref{acceptance}
gave $\epsilon\Delta\Omega = 0.66 \times 10^{-4}$
for hydrogen and $\epsilon\Delta\Omega = 0.90 \times 10^{-4}$
for deuterium. The factors $F = 0.90 \pm 0.09$ for hydrogen
and $F = 0.96 \pm 0.06$ for deuterium 
account for small differences
between the measured and the calculated acceptances
due to various detector efficiencies, {\it etc}.
The factor $c_{stop} = 0.85 \pm 0.01$ accounts
for the fraction of incident pions stopping in 
liquid hydrogen or liquid deuterium
and was computed via a Monte Carlo simulation \cite{Wr98}.
The factor $c_{bm} = 0.98$ for hydrogen and 
$c_{bm} = 0.91$ for deuterium accounts
for efficiencies of events
surviving the beam telescope amplitude 
and C-counter timing cuts 
that are omitted
from the simulation
(the uncertainties 
in $c_{bm}$ are negligible).

Using these values and Eqn.\ \ref{e:br}
we obtained absolute branching ratios 
of $( 3.02 \pm 0.27 (\mbox{stat.}) \pm 0.31  
(\mbox{syst.})) \times 10^{-5}$
for hydrogen
and $(1.42 \pm ^{0.09}_{0.12} (\mbox{stat.}) \pm 0.11  
(\mbox{syst.})) \times 10^{-5}$
for deuterium.
The corresponding values
of ratios $R ( 1\gamma / 2\gamma )$
of double radiative capture to single radiative capture
are $( 7.68 \pm 0.69 (\mbox{stat.}) \pm 0.79 (\mbox{syst.}))  \times 10^{-5}$ 
for hydrogen and
$( 5.44 \pm^{0.34}_{0.46} (\mbox{stat.}) 
\pm 0.42 (\mbox{syst.})) \times 10^{-5}$ 
for deuterium.
The quoted uncertainties are dominated by the
statistical error in the number of photon pairs
and the systematic error in the acceptance
for double radiative capture events. 

For completeness we also quote the partial widths $\Gamma$ 
for  double radiative capture. We use the recent measurements
of the total widths of the 1S atomic states 
of 0.86$\pm$0.07~$eV$ in pionic hydrogen 
(Schroeder {\it et al.}\ \cite{Sc99})  
and $1.13$$\pm$0.13~eV in pionic deuterium 
(Hauser {\it et al.}\ \cite{Ha98}).
Combined with our absolute branching ratios
they yield partial widths of $\Gamma_{1S} = 26.0 \pm 4.1 $~$\mu$eV 
for hydrogen and $\Gamma_{1S} = 16.2\pm 2.5$~$\mu$eV for deuterium. 

The sensitivity of the branching ratios to the opening
angle cut and the sum energy cut are given in
Table \ref{t: sensitivity}. Variations of the opening angle
cut from $-0.2 \geq y \geq 0.0$ changed the branching ratio
by 7\% in hydrogen and 2\% in deuterium.
Variation of sum energy minimum
from $60$ to $80$~MeV 
and sum energy maximum from $150$ to $170$~MeV
changed the branching ratios
by 5\% in hydrogen and 7\% in deuterium.
A separate analysis of the two hydrogen data-sets 
with the different level-three triggers 
(Sec.\ \ref{trigger}) gave branching ratios that 
differed by 7\%. In all cases the variations are
significantly smaller than the quoted errors on the branching ratios.

We stress that the quoted branching ratios
are for photon pairs with all possible angles 
and all possible energies.
Their derivation assumes the 
energy-angle distributions of photon
pairs calculated by Beder \cite{Be79}
-- the calculation and measurement being
consistent for the visible region of the 
angle spectra $y > -0.1$ 
and energy spectra $\omega > 25$~MeV.

To test the sensitivity 
to the energy-angle distribution
of the photon pairs
we also extracted branching ratios
with two other energy-angle distributions.
Assuming an energy-angle distribution given by 
only $\pi$$\pi$ annihilation graphs 
the resulting branching ratios are decreased
by 13\% in hydrogen and 14\% in deuterium,
the decrease reflecting
the larger proportion of photon pairs 
with small opening angles. 
Assuming an energy-angle distribution 
given by three-body $\gamma$$\gamma$n phase space
the resulting branching ratios are increased
by 5\% in hydrogen and 6\% in deuterium,
the increase reflecting
the smaller proportion of photon pairs 
with small opening angles. 

\subsection{Discussion of results for hydrogen}

Our measured absolute branching ratio $( 3.02 \pm 0.27 (\mbox{stat.}) \pm 0.31  
(\mbox{syst.})) \times 10^{-5}$ for double radiative
capture on hydrogen is in fair agreement with the
predicted values of 5.1$\times$10$^{-5}$ (Joseph \cite{Jo60}),
5.1$\times$10$^{-5}$ (Lapidus and Musakhanov \cite{La72})
and 5.4$\times$10$^{-5}$ (Beder \cite{Be79}) that
embrace the dominance of the $\pi \pi \rightarrow \gamma \gamma$
annihilation mechanism.
The measured branching ratio is definitely grossly under-estimated
by either an $\ell = 0$ NN mechanism or an $\ell = 0$ $\pi$N mechanism
and definitely grossly over-estimated by an $\ell = 1$ $\pi$N mechanism
(see Sec.\ \ref{dynamics} and Ref.\ \cite{Be79} for details).
Intriguingly, the measurement is very close to the prediction 
3.5$\times$10$^{-5}$ of the $O(1)$ $\pi$$\pi$ 
mechanism in the absence of the $O(1/M_n)$ NN correction 
(we offer no rationale for the omission of the $O(1/M_n)$ corrections
from the NN graphs).

\begin{figure}
\begin{center}
\epsfig{file=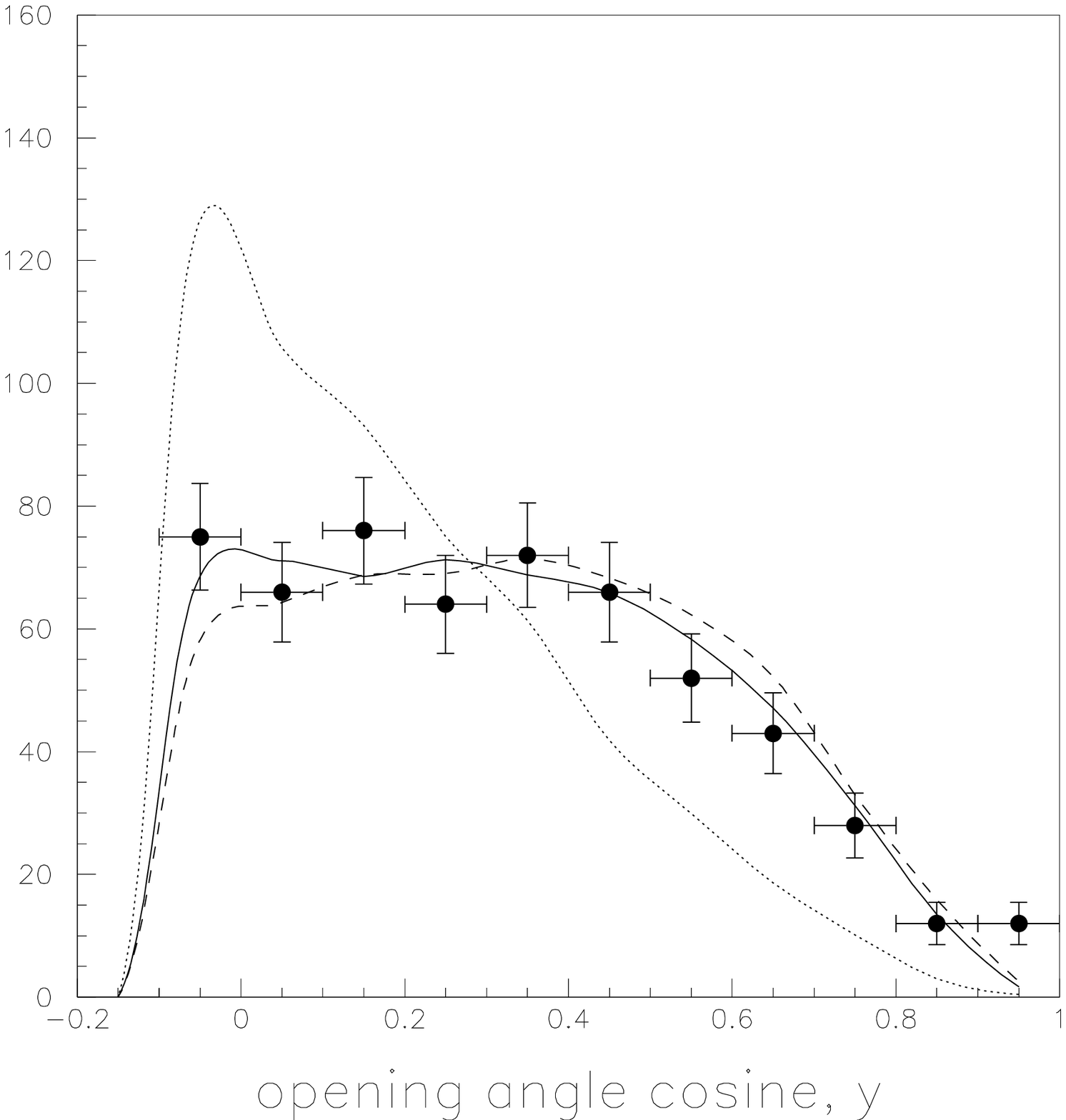,height=8.0cm}
\end{center}
\caption{Comparison of the opening angle distribution
from the experimental data (closed circles)
and the theoretical calculations (curves) for 
double radiative capture in pionic hydrogen. The solid
curve is the full calculation, the dashed curve
is the $\pi$$\pi$ term only, and the dotted curve
is the NN term only. Note the theoretical curves
have been normalized to the summed counts of the
experimental spectrum for the comparison of the 
angular dependences.}
\label{fig:1h_compare}
\end{figure}

A comparison between the theoretical predictions and the
experimental data for the two-photon opening angle
distribution is shown in Fig.\ \ref{fig:1h_compare}. 
Indicated are the results of the 
full $\ell = 0$  calculation, $O(1)$ $\pi$$\pi$ contribution 
only and $O(1/M_n)$ NN contribution only.
Note that the theoretical curves have been
convoluted with the response function of the 
detector system and incorporate such effects as
$e^{\pm}$ energy loss and multiple scattering
(see Sec.\ \ref{acceptance} for details). 
The measured spectra for opening angles $y > -0.1$
are reasonably consistent with the angular dependences
of both the full calculation and the
$O(1)$ $\pi$$\pi$ term only.
The experimental data are not consistent with the 
angular dependence of the NN mechanism which falls
too steeply with increasing $y$.

We consider the experimental results for the absolute
branching ratio and the opening angle
distribution to be in support of the underlying theoretical idea 
of a dominant $\pi$$\pi \rightarrow \gamma \gamma$ mechanism
in the $\ell = 0$ elementary process.
However, the measured branching ratio of $3.02 \pm 0.27 (\mbox{stat.}) \pm 0.31  
(\mbox{syst.}) \times 10^{-5}$ and calculated branching 
ratio of 5.1-5.4$\times$10$^{-5}$ do differ by roughly 40\%.
For radiative pion photoproduction 
the complete set of leading order (LO), next-to-leading order 
(NLO) and next-to-next-to-leading order (NNLO) graphs 
have been enumerated by Kao {\it et al.}\ \cite{Ka04}
within heavy-baryon chiral perturbation theory; however,
no such program has been performed 
for the double radiative process.
It is therefore possible that the remaining discrepancy 
between the measured branching ratio and the
calculated branching ratio may reflect the neglect 
of either certain NLO graphs or all NNLO graphs 
in the existing calculations of the double radiative process.
A modern calculation of double radiative capture
within heavy baryon chiral perturbation theory
would be welcome.

As mentioned in Sec.\ \ref{s-introduction}, at appropriate kinematics 
the $\pi$$\pi \rightarrow \gamma \gamma$ annihilation graph in double 
radiative capture and 
$\pi$$\gamma$$\rightarrow$$\pi$$\gamma$ 
scattering graph in radiative pion photoproduction
are potentially sensitive 
to the charged pion polarizability.
Indeed, a recent experiment at MAMI \cite{Ah05},
and an earlier experiment at Lebedev \cite{Ai86},
have attempted to extract the polarizability 
from the $\pi$$\gamma$$\rightarrow$$\pi$$\gamma$ 
contribution to the $\gamma p \rightarrow \gamma \pi^+ n$ reaction.
Although, for threshold double radiative capture
the pion polarizability effects are too small 
to be measurable,
the dominant contribution of the $\pi$$\pi$ graph
and vanishing contributions of the $\pi$N graphs 
do make the process an interesting yardstick 
for related theoretical studies
of radiative pion production.

\subsection{Discussion of results for deuterium}

In comparing our results for hydrogen and
deuterium we focus on the ratio 
$R ( 2 \gamma / 1 \gamma )$ and the two photon
kinematical distributions.
A quantitative comparison of 
absolute branching ratios is more difficult
and less meaningful
as pion absorption on nucleon pairs 
is important for deuterium but inapplicable for hydrogen.
  
Using our measured branching ratios for double radiative
capture and the published branching ratios for single radiative 
capture we derived values for 
$R ( 2 \gamma / 1 \gamma )$ of
$( 7.68 \pm 0.69 (\mbox{stat.}) \pm 0.79 (\mbox{syst.}))  \times 10^{-5}$ 
for hydrogen and
$( 5.44 \pm^{0.34}_{0.46} (\mbox{stat.})  \pm 0.42 (\mbox{syst.})) \times 10^{-5}$ 
for deuterium (Sec.\ \ref{br}). 
The results imply a decrease in $R ( 2 \gamma / 1 \gamma )$ 
from hydrogen to deuterium by about $(29 \pm 12)$\% 
(one small caveat 
is our detection of photon pairs is 
limited to $y > -0.1$ and $\omega > 20$~MeV).
In addition,
while the angular distributions are similar for hydrogen and deuterium,
the energy distributions show evidence of lower average photon energies
in deuterium (see Fig.\ \ref{fig:1h_2h_signal}).\footnote{Note that 
although the photon detection efficiencies
were significantly different in hydrogen and deuterium
their energy dependences were very similar (see Sec.\ \ref{acceptance}) for details)}

The decrease in $R ( 2 \gamma / 1 \gamma )$
and the change in photon energies from
hydrogen to deuterium
are probably not surprising.
Single radiative capture and double radiative capture 
involve different reaction mechanisms
and different momentum transfer, 
and therefore effects such as Pauli blocking, 
initial-state wavefunctions, final-state interactions, 
{\it etc.}, may be different for these two processes.
For example, the lower photon energies 
in deuterium capture compared to hydrogen capture 
may perhaps reflect the fact that energy is imparted to the spectator
neutron through the final-state interactions 
in the deuterium process.
In addition, the smaller $R ( 2 \gamma / 1 \gamma )$ ratio
in deuterium capture compared to hydrogen capture 
may perhaps reflect a greater effect from Pauli blocking
at the smaller momentum transfer of the double radiative process.
Of course, to properly address 
the hydrogen-deuterium comparison
a detailed calculation of deuterium capture is needed. 

\subsection{Comparison to nuclear double radiative capture.}

The world data for double radiative capture
on complex nuclei consist of
results for $^{9}$Be and $^{12}$C 
by Deutsch {\it et al.}\ \cite{De79} at CERN
and results for $^{12}$C from 
Mazzucato {\it et al.}\ \cite{Ma80} at TRIUMF.
The  experiment of Deutsch {\it et al.}\ obtained a partial branching
ratio of $(1.0 \pm 0.1 ) \times 10^{-5}$
for $^{9}$Be and  $(1.4 \pm 0.2 ) \times 10^{-5}$
for $^{12}$C for photon energies exceeding $25$~MeV.
The experiment of  Mazzucato {\it et al.}\ obtained a partial branching
ratio of $(1.2 \pm 0.2 ) \times 10^{-5}$
for $^{12}$C for photon energies exceeding $17$~MeV.
Both experiments reported an increased yield
of photon pairs at large opening angles
although Mazzucato {\it et al.}\ have suggested
that target contamination by hydrogenous materials
may perhaps be responsible. The Mazzucato {\it et al.}\ experiment 
also reported a preference 
for unequal partition of photon energies.

In comparing the nuclear $(\pi, 2\gamma)$ data to
our $(\pi, 2\gamma)$ data for hydrogen and deuterium we consider the
ratio $R ( 1 \gamma / 2 \gamma )$ in order to avoid
the complication of accounting for 
non-radiative processes in complex nuclei. For $^{12}$C,
we derived values for  
$R ( 2 \gamma / 1 \gamma )$ of $( 7.2 \pm 1.3 ) \times 10^{-4}$ 
from Mazzucato {\it et al.}\
and $( 8.4 \pm 1.3 ) \times 10^{-4}$ 
from Deutsch {\it et al.}\
when normalized using the branching ratio for single radiative capture
from Bistirlich {\it et al.}\ \cite{Bi72}.
For $^{9}$Be, we obtained a value for 
$R ( 2 \gamma / 1 \gamma )$ of  $( 4.8 \pm 1.3 ) \times 10^{-4}$ 
when normalized using the mean branching ratio
for single radiative capture from the 
neighboring $^7$Li-$^{10}$B nuclei \cite{Ba77}. 
Clearly -- as summarized in Table \ref{t:rvalues} -- 
the values of $R ( 2 \gamma / 1 \gamma )$ 
on light nuclei are much larger
than the corresponding ratios for hydrogen
$( 7.68 \pm 0.69 \pm 0.79 ) \times 10^{-5}$ 
and deuterium $( 5.44 \pm^{0.34}_{0.46} (\mbox{stat.}) 
\pm 0.42 (\mbox{syst.}) \times 10^{-5}$.
This increase in $R ( 2 \gamma / 1 \gamma )$ from 
$Z = 1$ to light nuclei was anticipated
by several authors \cite{Be79,Ch79}
as an evolution from predominantly annihilation dynamics 
in S-state capture to predominantly bremsstrahlung dynamics
in atomic P states (see Sec.\ \ref{dynamics} for details). For instance, an 
increase in $R( 2\gamma / 1 \gamma )$ from $\ell = 0$ capture  
to $\ell = 1$ capture of approximately ten-fold
was predicted by Beder in Ref.\ \cite{Be79}.

Another difference between our data for
$Z = 1$  and earlier data for
complex nuclei is the photon energy spectra. 
For $Z = 1$ the energy spectra (Fig.\ \ref{fig:1h_2h_signal}) 
indicate a preference for the equal partition of the photon energies. 
For  $^{12}$C the energy spectra (Refs.\ \cite{De79} and \cite{Ma80}) 
indicate a preference for the unequal partition of the photon energies. 
This difference between $\ell = 0$ capture and $\ell = 1$ capture 
was also anticipated in early theoretical studies of double radiative capture 
by Beder \cite{Be79} and Christillin and Ericson \cite{Ch79}, 
and likewise reflects the evolving dynamics from annihilation in S-state capture 
to bremsstrahlung in P-state capture.

Another observation of experiments on nuclei
was a backward peak in the angular distribution
of the photon pairs.
This feature was not predicted by the earlier 
calculation of Christillin and Ericson \cite{Ch79} 
but was obtained in the later calculation of Gil and Oset \cite{Gi95}.
Unfortunately, the intense background of back-to-back photon
pairs from $\pi^o \rightarrow \gamma \gamma$ decay
made unthinkable the measurement of $Z =1$ double radiative capture 
at large opening angles.

In summary, we claim the large difference in 
$R( 2\gamma / 1 \gamma )$ between the elementary 
and nuclear processes, and significant difference 
in the energy partition between the elementary and 
nuclear processes, are clear evidence of 
the evolving dynamics of the double radiative process
from S-state to P-state capture. 
This conclusion implies that earlier arguments
for the sensitivity of the $(\pi^- , 2\gamma )$ nuclear reaction
to the pion field in the nuclear medium,
that were founded on the dominance of the
$\pi\pi \rightarrow \gamma\gamma$ mechanism, 
are likely not justified.

\section{Conclusion}
\label{conclusions}

We have performed the first measurements 
of double radiative capture 
on pionic hydrogen and pionic deuterium.
The measurements were performed 
using the RMC spectrometer 
at the TRIUMF cyclotron 
by recording photon pairs 
from pion stops 
in liquid hydrogen and deuterium targets.
We have determined the absolute branching ratios 
for double radiative capture, 
the ratios $R ( 2\gamma / 1\gamma)$
of double radiative capture
to single radiative capture,
and the energy-angle distributions
of the resulting photon pairs.

For hydrogen we obtained an absolute branching ratio 
of $( 3.02 \pm 0.27 (\mbox{stat.}) \pm 0.31  
(\mbox{syst.}) ) \times 10^{-5}$
and ratio $R ( 2 \gamma / 1 \gamma )$ of 
$( 7.68 \pm0.69(stat) \pm 0.79(syst)) \times 10^{-5}$.
The measured branching ratios and energy-angle distributions
are in fair agreement with the theoretical predictions
of Joseph \cite{Jo60}, Lapidus and Musakanov \cite{La72} 
and Beder \cite{Be79} and lend support
to the theoretical claim of a 
$\pi$$\pi$$\rightarrow$$\gamma$$\gamma$ annihilation mechanism.
This mechanism implies
that double radiative capture
from  $\ell = 0$ orbitals inherits 
an intriguing sensitivity 
to the nucleon's pion cloud
and the $\pi\pi \rightarrow \gamma \gamma$ vertex.

For hydrogen the measured and 
calculated branching ratio
do, however, differ by approximately 40\%. 
One possible explanation is the neglect
of either certain NLO graphs
or all NNLO graphs in the existing calculations
of the double radiative process. 
Given the relationship between
double radiative pion capture $\pi^- p \rightarrow \gamma \gamma n$ 
and radiative pion photoproduction $\gamma  p \rightarrow \gamma \pi^+ n$
-- and their relevance to the determination
of the pion's polarizability --
we believe that resolving this discrepancy
is of some importance.
A modern calculation of double radiative pion capture 
within heavy-baryon chiral perturbation 
theory would be welcome.

For deuterium we obtained an absolute branching ratio of
$( 1.42 \pm ^{0.09}_{0.12} (\mbox{stat.}) \pm 0.11  
(\mbox{syst.}) ) \times 10^{-5}$
and ratio $R ( 2 \gamma / 1 \gamma )$ of
$( 5.44 \pm ^{0.34}_{0.46}(stat) \pm0.42(syst)) \times 10^{-5}$.
We found that the deuterium ratio $R ( 2 \gamma / 1 \gamma )$
is smaller than the hydrogen ratio $R ( 2 \gamma / 1 \gamma )$
by $(29 \pm 12)$\%. 
We also found the photon energies 
are shifted somewhat to lower energies 
in the deuterium data compared to the hydrogen data.
The decrease in $R ( 2 \gamma / 1 \gamma )$
and shift in photon energies are possibly revealing
the spectator neutron's role 
via final-state interactions and Pauli blocking effects.

Lastly, we compared
our results for hydrogen and deuterium with the earlier results
for beryllium and carbon. The relative branching
ratios $R ( 2 \gamma / 1 \gamma )$ for light
nuclei are nearly ten times larger 
than the corresponding values
for the hydrogen isotopes.
In addition, the energy distributions are markedly 
different for the light nuclei and the hydrogen isotopes;
the former favoring unequal energy partition with 
the latter favoring equal energy partition.
These observations are consistent with the dynamics evolving
from a $\pi\pi \rightarrow \gamma\gamma$ annihilation mechanism
for $\ell = 0$ capture to a $\pi$N bremsstrahlung mechanism 
for $\ell = 1$ capture.
This implies that nuclear double radiative pion capture 
may not inherit
the high sensitivity 
to the pion field
in the nuclear medium
as was originally proposed.

\section{Acknowledgements}
\label{acknowledgements}

We would like to thank the staff of the TRIUMF laboratory
for the operation of the cyclotron and the
maintenance of the RMC spectrometer.
We would also like to thank Dr. Ren\'{e}e Poutissou for
assistance with the data acquisition and
Dr. Dennis Healey for assistance with the cyrogenic targets.
We also thank Drs.\ Fearing, Kao, Oset and Wilkin 
for valuable communications on the theoretical aspects
of the double radiative process.
Lastly, we are indebted
the National Science Foundation (USA),
Natural Sciences and Engineering Research Council (Canada),
and Jeffress Memorial Trust and
the William \& Mary Endowment
for their financial support.

%
%

\newpage

\begin{table*}
\caption{Compilation of previously published branching ratios 
for negative pion capture on hydrogen and deuterium.}
\label{t: H BR}
\begin{center}
\begin{tabular}{ccc}
  & & \\
capture mode & expt.\ & branching ratio \\
  & & \\ 
\hline
  & & \\ 
$\pi^- p \rightarrow \pi^o n$   &  Spuller {\it et al.}\ \cite{Sp77}  & $0.607 \pm 0.004$ \\
$\pi^- p \rightarrow \gamma n$ &  Spuller {\it et al.}\ \cite{Sp77}  & $0.393 \pm 0.003$ \\
$\pi^- p \rightarrow e^+ e^- n$ &  Samios \cite{Sa61}  & $(2.72 \pm 0.19 )$$\times$$10^{-3}$ \\
$\pi^- d \rightarrow n n$        &  Highland {\it et al.}\ \cite{Hi81}  & $0.739 \pm 0.010$ \\
$\pi^- d \rightarrow \gamma n n$ &  Highland {\it et al.}\ \cite{Hi81}  & $0.261 \pm 0.004$ \\
$\pi^- d \rightarrow \pi^o n n$    &  MacDonald {\it et al.}\ \cite{Ma77} & 
$( 1.45\pm0.19 )$$\times$$10^{-4}$ \\
 & & \\
\end{tabular}
\end{center}
\end{table*}

\begin{table*}
\caption{Summary of tracking cuts for positrons and electrons
and ``nonwrap arounds'' and ``wrap arounds''. The upper rows
are the hydrogen values
and the lower rows are the deuterium values.}
\label{t: 1H tracking cuts}
\begin{center}
\begin{tabular}{ccccc}
  & & & & \\
track    & nonwrap & wrap  & nonwrap & wrap   \\
parameter & $e^+$   & $e^+$ & $e^-$   & $e^-$  \\
  & & & & \\ 
\hline
  & & & & \\ 
\underline{hydrogen cuts} & & & & \\ 
  & & & & \\  
$S_{xy}^2 ( cm^2 )$ & $<0.0048$ & $<0.0270$ & $<0.0038$ & $<0.0210$ \\ 
$S_{z}^2 ( cm^2 )$  &  $<0.2$ & $<3.0$ & $<0.2$ & $<3.0$ \\
$N_{xy}$   & $\geq 9$ & $\geq 9$ & $\geq 9$ & $\geq 9$ \\
$N_{z}$    & $\geq 3$ & $\geq 3$ & $\geq 3$ & $\geq 3$ \\
$d_{wc} ( cm )$ & $<1.0$ & $<1.0$ & $<1.0$ & $<1.0$ \\
  & & & & \\ 
\underline{deuterium cuts} & & & & \\ 
  & & & & \\ 
$S_{xy}^2 ( cm^2 )$ & $<0.0048$ & $<0.0270$ & $<0.0038$ & $<0.0210$ \\ 
$S_{z}^2 ( cm^2 )$  &  $<0.2$ & $<3.0$ & $<0.2$ & $<3.0$ \\
$N_{xy}$   & $\geq 9$ & $\geq 9$ & $\geq 9$ & $\geq 9$ \\
$N_{z}$    & $\geq 3$ & $\geq 3$ & $\geq 3$ & $\geq 3$ \\
$d_{wc} ( cm )$ & - & - & - & - \\
  & & & & \\ 
\end{tabular}
\end{center}
\end{table*}

\begin{table*}
\caption{Summary of the photon cuts for 
``non-wrap around'' photons ({\it i.e.}\ containing
no wrap-around tracks)  and ``wrap around'' photons 
({\it i.e.}\ containing wrap-around tracks). The upper rows
are the hydrogen values 
and the lower rows are the deuterium values.}
\label{t: 1H photon cuts}
\begin{center}
\begin{tabular}{ccc}
  & & \\
photon     & nonwrap & wrap  \\
parameter &   & \\
  & & \\ 
\hline
  & & \\ 
\underline{hydrogen cuts} & & \\ 
 & & \\
$d_{xy} ( cm )$      &  $<6.0$     & $<7.0$ \\ 
$d_{z} ( cm )$       &  $<6.0$     & $<7.0$ \\
$r_{close} ( cm )$   & $< 8$    & $< 8$ \\
$z_{close} ( cm )$   & $-12 < z < +14$ & $-12 < z < +14$ \\
  & & \\ 
\underline{deuterium cuts} & & \\ 
  & & \\ 
$d_{xy} ( cm )$      &  $<6.0$     & $<7.0$ \\ 
$d_{z} ( cm )$       &  $<12.0$     & $<14.0$ \\
$r_{close} ( cm )$   & $< 8$    & $< 8$ \\
$z_{close} ( cm )$   & $-25 < z < +27$ & $-25 < z < +27$ \\
  & & \\ 
\end{tabular}
\end{center}
\end{table*}

\begin{table*}
\caption{Summary of background estimates and 
background limits for the hydrogen analysis 
and the deuterium analysis. The first row is the
two-photon events with summed energies 80-150~MeV before
the background subtraction and the last low is
two-photon events with summed energies 80-150~MeV after
the background subtraction.}
\label{t: counts}
\begin{center}
\begin{tabular}{ccc}
 & & \\
  background source & H$_2$ & D$_2$  \\
 & & \\
\hline
 & & \\ 
$80$-$150$~MeV events before sub.\ & 566 & 327 \\
 & & \\ 
 $\pi^o$ bkd.\  & 53$\pm$30 & 2$\pm$1 \\
 $\pi$$\pi$ bkd.\  & 40$\pm$8 & 20$\pm$2 \\
 $\mu$$\pi$ bkd.\ & 0 & 15$\pm$4 \\
 Dalitz bkd.\ & $<$2 & $<$1 \\
 nucl.\ $(\pi , 2\gamma )$ & $<$10 & $<$20 \\
 & & \\
$80$-$150$~MeV events after sub.\ & 465$\pm$$^{39}_{40}$ & 292$\pm$$^{19}_{27}$ \\
 & & \\ 
\end{tabular}
\end{center}
\end{table*}

\begin{table*}
\caption{Summary of quantities for the calculations
of the total branching ratios of double radiative capture
on pionic hydrogen and pionic deuterium.  Note that the 
quoted acceptances assumes the energy-angle distributions 
of Beder.}
\label{t: 1H 2pi bkd}
\begin{center}
\begin{tabular}{ccc}
  & & \\
 quantity & H$_2$  & D$_2$  \\
 & & \\
\hline
 & & \\ 
N$_{\gamma \gamma}$ & 465$\pm$$^{39}_{40}$ & 292$\pm$$^{19}_{27}$ \\
N$_{\pi^-}$ & (3.10$\pm$0.03)$\times$10$^{11}$ &  (3.07$\pm$0.15)$\times$10$^{11}$  \\
$\epsilon$$\Delta$$\Omega$ & 0.66$\times$10$^{-4}$ & 0.90$\times$10$^{-4}$ \\
F & 0.90$\pm$0.09 & 0.96$\pm$0.06 \\
C$_{stop}$ &         0.85$\pm$0.01 &           0.85$\pm$0.01 \\
C$_{bm}$ & 0.98 & 0.91 \\
 & & \\ 
absolute  B.R.\ & 3.02$\pm$0.27(stat)$\pm$0.31(syst)$\times$10$^{-5}$ & 
1.42$\pm$$^{0.09}_{0.12}$(stat)$\pm$0.11(syst)$\times$10$^{-5}$ \\ 
 & & \\ 
ratio R(2$\gamma$/1$\gamma$) & 7.68$\pm$0.69(stat)$\pm$0.79(syst)$\times$10$^{-5}$ & 
5.44$\pm$$^{0.34}_{0.46}$(stat)$\pm$0.42(syst)$\times$10$^{-5}$ \\
 & & \\ 
\end{tabular}
\end{center}
\end{table*}

\begin{table*}
\caption{Sensitivity of the absolute branching ratio to the analysis parameters, different
data-sets and the theoretical energy-angle distributions for the hydrogen
analysis and the deuterium analysis. We have combined in quadrature the 
various statistical and systematic uncertainties.}
\label{t: sensitivity}
\begin{center}
\begin{tabular}{ccc}
 & & \\
  cuts/data-set/theory & H$_2$ & D$_2$  \\
 & & \\
\hline
 & & \\ 
$80 < \omega_1 + \omega_2 < 150$,  $y >  0.0 $  & $( 2.94 \pm 0.41 ) \times 10^{-5}$ & $(1.39\pm^{0.14}_{0.16}  ) \times 10^{-5}$ \\
$80 < \omega_1 + \omega_2 < 150$,  $y > -0.1 $  & $( 3.02 \pm 0.41 ) \times 10^{-5}$ & $(1.42\pm^{0.14}_{0.16}  ) \times 10^{-5}$ \\
$80 < \omega_1 + \omega_2 < 150$,  $y > -0.2 $  & $( 3.14 \pm 0.41 ) \times 10^{-5}$ & $(1.39\pm^{0.14}_{0.16}  ) \times 10^{-5}$ \\
 & & \\
$70 < \omega_1 + \omega_2 < 150$,  $y >  -0.1 $  & $( 3.11 \pm 0.41 ) \times 10^{-5}$ & $( 1.48\pm^{0.14}_{0.17} ) \times 10^{-5}$  \\
$60 < \omega_1 + \omega_2 < 150$,  $y >  -0.1 $  & $( 3.17 \pm 0.41 ) \times 10^{-5}$ & $( 1.48\pm^{0.14}_{0.17} ) \times 10^{-5}$  \\
 & & \\
$80 < \omega_1 + \omega_2 < 160$,  $y >  -0.1 $  & $( 3.11 \pm 0.41 ) \times 10^{-5}$ & $( 1.42\pm^{0.14}_{0.16} ) \times 10^{-5}$ \\
$80 < \omega_1 + \omega_2 < 170$,  $y >  -0.1 $  & $( 3.05 \pm 0.41 ) \times 10^{-5}$ & $( 1.38\pm^{0.14}_{0.16} ) \times 10^{-5}$ \\
 & & \\
hydrogen data-set I  & $( 2.94 \pm 0.49 ) \times 10^{-5}$ & n/a \\
hydrogen data-set II & $( 3.15 \pm 0.49 ) \times 10^{-5}$ & n/a \\
 & & \\ 
Beder, all graphs & $( 3.02 \pm 0.41 ) \times 10^{-5}$ & $(1.42\pm^{0.14}_{0.16} ) \times 10^{-5}$ \\
Beder, $\pi$$\pi$ graphs & $( 2.63 \pm 0.41 ) \times 10^{-5}$ & $( 1.22 \pm^{0.12}_{0.14} ) \times 10^{-5}$ \\
phase space & $( 3.16 \pm 0.41 ) \times 10^{-5}$ & n/a \\ 
 & & \\ 
\end{tabular}
\end{center}
\end{table*}

\begin{table*}
\caption{Summary of determinations of ratios $R(2\gamma / 1\gamma )$
from our hydrogen and deuterium data and earlier beryllium and
carbon data. Note that for beryllium and
carbon the ratios are for photon energies 
$>$25~MeV in Deutsch {\it et al.}\ and
$>$7~MeV Mazzucato {\it et al.}.}
\label{t:rvalues}
\begin{center}
\begin{tabular}{ccc}
 & & \\
 target & Ref.\ & $R(2\gamma / 1\gamma )$ \\
 & & \\
\hline
& & \\
$^{1}$H & this work & $( 7.68 \pm 0.69 \pm 0.79 ) \times 10^{-5}$      \\
$^{2}$H & this work & $( 5.44 \pm \pm^{0.34}_{0.46} (\mbox{stat.}) 
                      \pm 0.42 (\mbox{syst.}) \times 10^{-5}$          \\
$^{9}$Be & Deutsch {\it et al.}\ \cite{De79}   & $( 4.8 \pm 1.3 ) \times 10^{-4}$ \\
$^{12}$C & Deutsch {\it et al.}\  \cite{De79}  & $( 8.4 \pm 1.3 ) \times 10^{-4}$ \\
$^{12}$C & Mazzucato {\it et al.}\ \cite{Ma80}  & $( 7.2 \pm 1.3 ) \times 10^{-4}$ \\
& & \\
\end{tabular}
\end{center}
\end{table*}

%
%

\newpage

\begin{figure}
\begin{center}
\epsfig{file=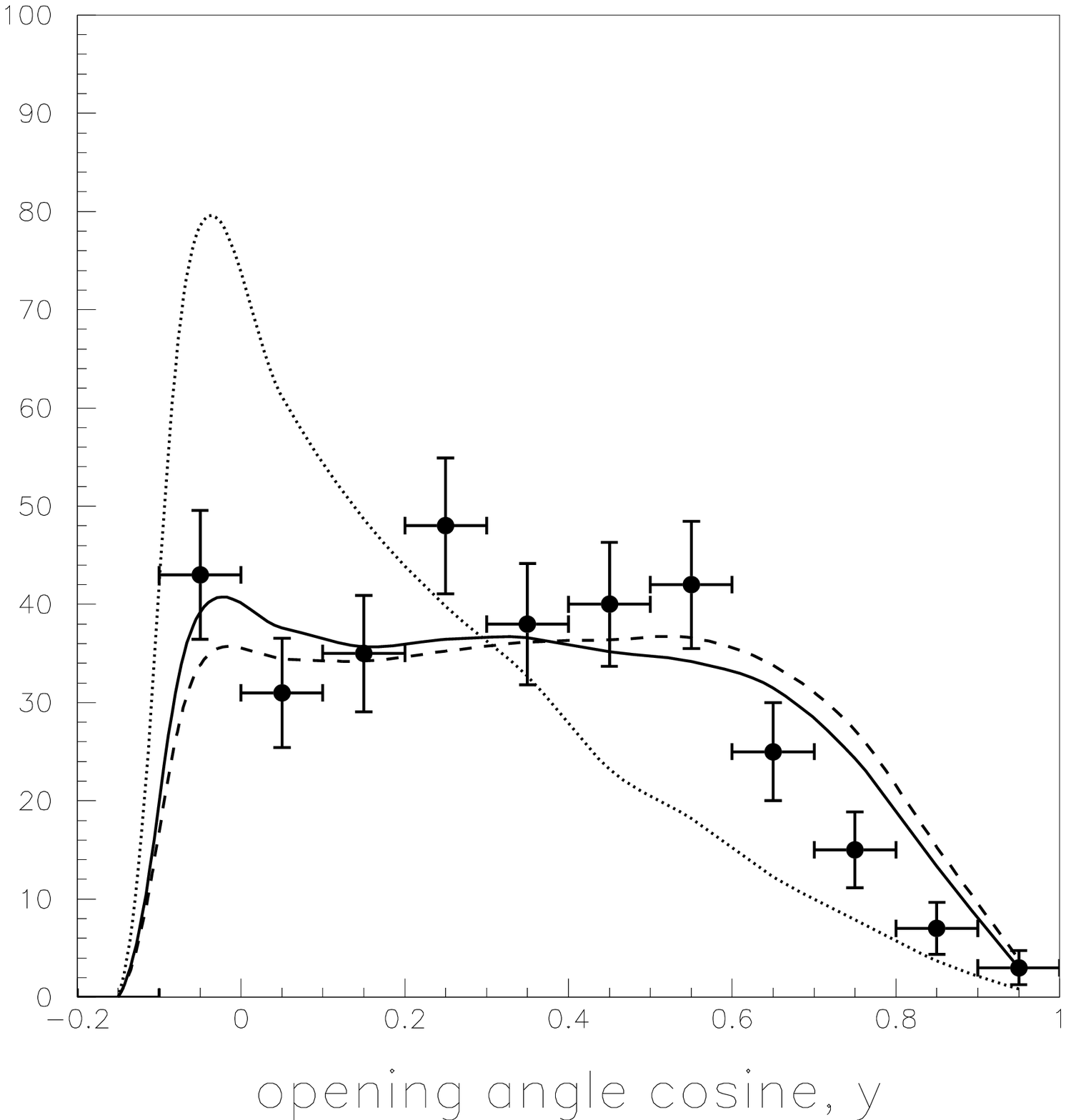,height=8.0cm}
\end{center}
\caption{Comparison of the opening angle distribution
from the experimental data (closed circles)
and the theoretical calculations (curves) for 
double radiative capture in pionic deuterium. The solid
curve is the full calculation, the dashed curve
is the $\pi$$\pi$ term only, and the dotted curve
is the NN term only. Note the theoretical curves
have been normalized to the summed counts of the
experimental spectrum for the comparison of the 
angular dependences.}
\label{fig:2h_compare}
\end{figure}

\end{document}